\documentclass[aps,prd,preprint,tightenlines,nofootinbib]{revtex4}
\usepackage{graphicx}
\include{rlFig}
\begin{document}



\title{Experimental Probes of Radio Wave Propagation near Dielectric Boundaries and Implications for Neutrino Detection}

\author{R. Alvarez}
\affiliation{University of Kansas, Lawrence, KS  66045}
\author{D.Z. Besson}
\affiliation{National Research Nuclear University MEPhI (Moscow Engineering Physics Institute), Moscow 115409 Russia}
\author{I. Bikov}
\affiliation{National Research Nuclear University MEPhI (Moscow Engineering Physics Institute), Moscow 115409 Russia}
\author{J.C. Hanson}
\affiliation{University of Kansas, Lawrence, KS  66045}
\author{A.M. Johannesen}
\affiliation{University of Kansas, Lawrence, KS  66045}
\author{J. Macy}
\affiliation{University of Kansas, Lawrence, KS  66045}
\author{S. Prohira}
\affiliation{University of Kansas, Lawrence, KS  66045}
\author{J. Stockham}
\affiliation{University of Kansas, Lawrence, KS  66045}
\author{M. Stockham}
\affiliation{University of Kansas, Lawrence, KS  66045}
\author{Al. Zheng}
\affiliation{University of Kansas, Lawrence, KS  66045}
\author{Am. Zheng}
\affiliation{University of Kansas, Lawrence, KS  66045}


\date{\today}

\begin{abstract}
Experimental efforts to measure neutrinos by radio-frequency (RF) signals resulting from neutrino interactions in-ice have intensified over the last decade. Recent calculations indicate that one may dramatically improve the sensitivity of ultra-high energy (``UHE''; $\ge 10^{18}$ eV) neutrino experiments via detection of radio waves trapped along the air-ice surface. Detectors designed to observe the ``Askaryan effect'' currently search for RF electromagnetic pulses propagating through bulk ice, and could therefore gain sensitivity if signals are confined to the ice-air boundary. To test the feasibilty of this scenario, measurements of the complex radio-frequency properties of several air-dielectric interfaces were performed for a variety of materials. Two-dimensional surfaces of granulated fused silica (sand), both in the lab as well as occurring naturally, water doped with varying concentrations of salt, natural rock salt formations, granulated salt and ice itself were studied, both in North America and also Antarctica. In no experiment do we observe unambiguous surface wave propagation, as would be evidenced by signals traveling with reduced signal loss and/or superluminal velocities, compared to conventional EM wave propagation. We therefore conclude that the prospects for experimental realization of such detectors are not promising.
\end{abstract}

\maketitle

\section{Introduction}
Cosmic rays propagating towards Earth from large redshifts can produce neutrinos by interacting with the cosmic microwave background\cite{GZK1,GZK2,GZK3}. The Aksaryan effect\cite{Askaryan} provides an attractive neutrino detection scheme whereby the charged and neutral current neutrino-nucleon interactions in a dielectric material create hadronic and electromagnetic cascades that radiate coherently in the $10^2-10^3$ MHz bandwidth, or wavelengths in the 30-300 cm range\cite{ZHS}.  Given its natural abundance in pure form, ice is the most convenient dielectric material, because the long radio attenuation length ($\approx$2 km in the coldest ice) facilitates observation of in-ice neutrino interactions by receivers (Rx) several km distant\cite{Barrella,hansJGlac}.


The Antarctic ice sheet represents a unique laboratory for studies of surface effects, given the dielectric contrast at the air/snow interface, and the semi-infinite character of the ice sheet. Antartic ice is a remote and largely inaccessible laboratory, however, so we have conducted a series of separate, and somewhat overlapping laboratory measurements to search for evidence of surface excitations. Inherent in each technique are systematic uncertainties, usually due to the finite scale of the measurement apparatus; redundancy is therefore essential in order to derive a composite picture. The techniques can be classified as follows:
\begin{itemize}
\item Measurement of signal amplitude reduction as a function of distance from a transmitter (Tx) near a dielectric boundary.
\item Direct measurements of group velocities for waves propagating near a dielectric boundary using signal impulses (in which case direct arrival times are determined), and corresponding determination of the index-of-refraction.
\item Measurements of phase velocity, by tracking the phase of a continuous-wave (CW) source propagating near that boundary, 
\item Inference of index-of-refraction by measurements of the Voltage Standing Wave Ratio (VSWR) of antennas near a dielectric boundary.  
\end{itemize}

We now discuss some of the studies that have been performed thus far. Most relevant to the goals of neutrino detection are two unpublished experiments performed by the ARIANNA vs. ARA collaborations in Antarctica -- the former was consistent with the observation of surface wave excitations connecting an in-ice transmitter at a depth of 20 m to a surface receiver, over a horizontal distance of 1000 m; the latter was similar geometrically, but with no signal detected at a depth of 25 m from a surface transmitter over a horizontal distance of 100 m. Future work in this direction is essential in determining the viability of this approach.

\section{Measurement of Attenuation Length Dependence on Distance}
The attenuation length is typically defined with respect to the electric field (directly proportional to the voltage measured at the feed point of our receiver dipole antennas).  In terms of power transmission, the (modified) Friis transmission equation states that the power received at the receiver after the signal has propagated a distance $d$ through the medium with field attenuation length $L$ is 

\begin{equation}
	P_{rec}/P_{trans} \sim {G_{Rx}G_{Tx} \over d^2} \exp{\left ( -2 {d \over L} \right )}.
	\label{P}
\end{equation}

Here, $G$ is a gain factor depending on the angular orientation of the transmitter and receiver.  For two aligned dipoles, $G$ is effectively of order unity.  Applying equation \ref{P} to two receivers at distances $r_1$ and $r_2$ from the transmitter, we have (in a three-dimensional material), for the voltages $V_1$ and $V_2$ measured at the two points:

\begin{equation}
	V_{2}/V_{1} = {r_1 \over r_2} \exp{\left ( -(r_2-r_1)/L \right )}
	\label{eqn:V}
\end{equation}

Equation \ref{eqn:V} assumes that $P \propto V^2$, and that the antennas are linear devices with a constant effective height converting electric fields into voltages (a standard property of dipole antennas, e.g.).  Comparing an antenna measurement in air (with no appreciable attenuation) to the dielectric material, equation \ref{eqn:V} becomes

\begin{equation}
	V_{diel}/V_{air} = {r_{air} \over r_{diel}} \exp{\left ( -(r_{diel}-r_{air})/L \right )}
	\label{Vair}
\end{equation}

In our case, we seek to investigate `trapping' of electric field flux within a boundary surface layer. In such a case, the dependence of signal power with distance falls more slowly than for flux spreading into the bulk ($1/r$ vs. $1/r^2$), suggesting significantly more advantageous neutrino detection efficiency for near-surface radio receivers rather than englacial radio receivers. 

\subsection{Laboratory Amplitude Measurements in Sand}
\begin{figure}
\begin{centering}
	\includegraphics[width=0.6\textwidth,angle=-90]{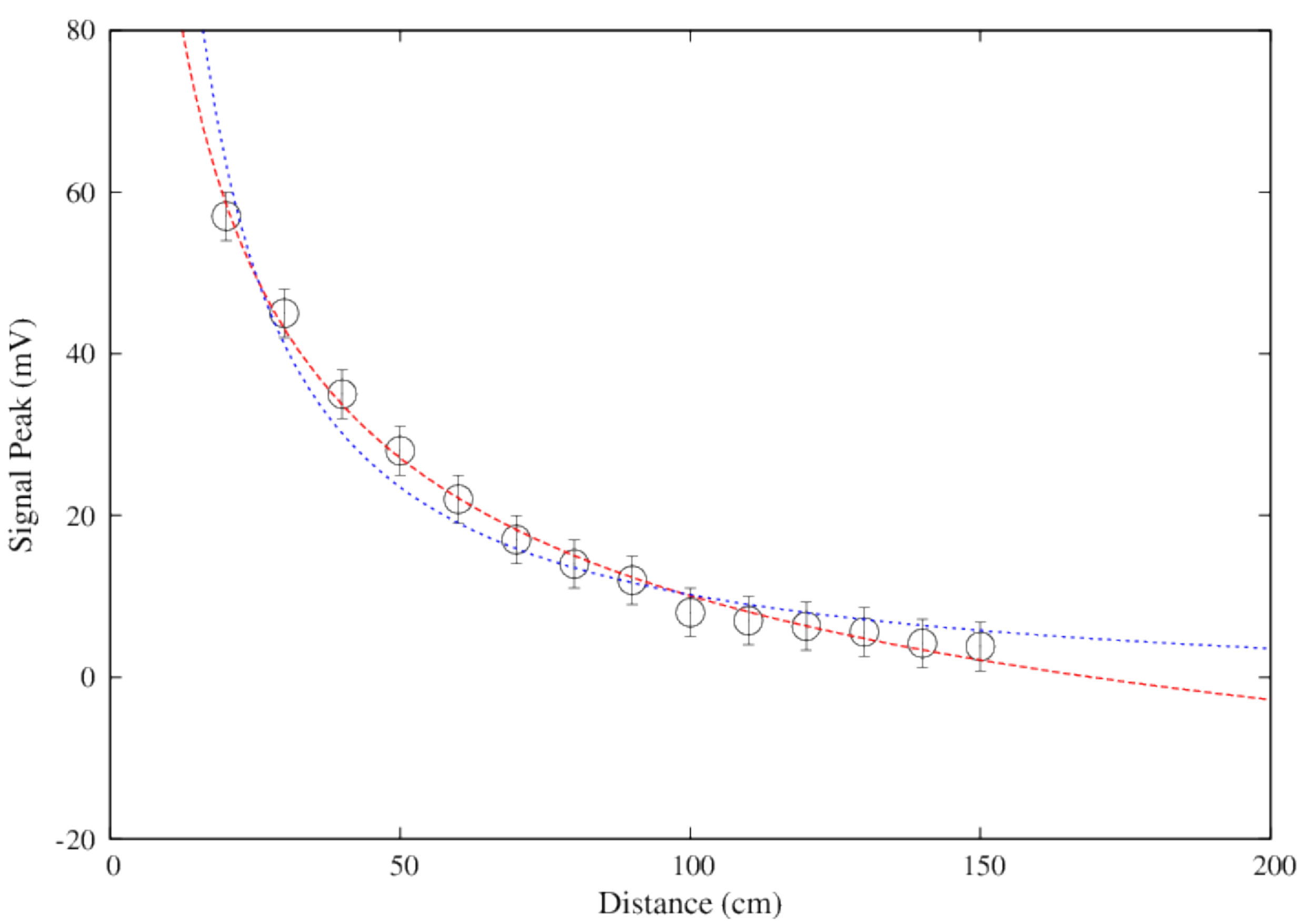}
	\caption{\it Near-surface amplitude measurements made in sand volleyball courts at University of Kansas. Data points shown as circles overlaid with fits to $1/r$ (red) vs. $1/\sqrt{r}$ (blue) functional forms. Errors represent error on average value displayed in each bin, and are therefore proportional to the square root of the inverse of the number of trials. \label{fig:SandAmplitude}}
\end{centering}
\end{figure}

On the University of Kansas campus, we have attempted to discern surface wave effects using the sand volleyball courts located at the University's gym and recreational facilities. These have the advantage of close proximity to the KU Physics Department, but the disadvantage of limited depth (typically 30 cm of dry sand, followed by a deeper layer susceptible to increasing moisture). Our measurements consisted of a series of peak voltages for impulses, measured as a function of separation distance between transmitter and receiver; for these measurements, both transmitter and receiver were half-buried in the sand. 
Each of the 5 cm diameter, 30 cm long RICE\cite{RICE} neutrino detection experiment's  
half-wave ``fat'' dipole antennas offers
good reception over the range 0.2--1 GHz.
The peak response of the antenna is measured to be at $\sim$450 MHz in air
(or 450 MHz/$n$ in a medium characterized by index-of-refraction $n$)
with a fractional bandwidth ${\Delta f\over f}\sim$0.2.
Figure \ref{fig:SandAmplitude} shows the result of this exercise. Although the data quality are reasonably good and favor the $1/r$ fit, we cannot conclusively discriminate between $V(r)\propto 1/r$ vs. $V(r)\propto 1/\sqrt{r}$, corresponding to spherical vs. planar flux-spreading, respectively.

\subsection{Amplitude Measurements at South Pole with the RICE Experiment}
As an alternative,
we have attempted to quantify planar signal flux trapping directly from data taken using a fat dipole antenna transmitter/receiver pair lowered into two neighboring iceholes drilled for the RICE experiment at the South Pole in 1998. Each of these mechanically holes is approximately 180 m deep and 12 cm in diameter.

\begin{figure}[ht]
\begin{minipage}{3.1in}
\includegraphics[width=\textwidth,angle=0]{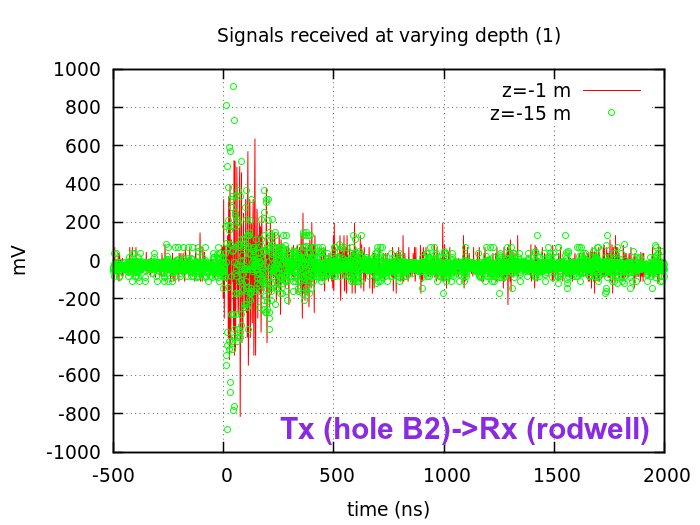}\caption{\it RICE Tx$\to$Rx waveforms obtained at varying depths (1m or 15m) for both Tx and Rx.} \label{fig:rodwell_X_B2}
\end{minipage}
\hspace{0.3in}
\begin{minipage}{3.1in}
\includegraphics[width=\textwidth,angle=0]{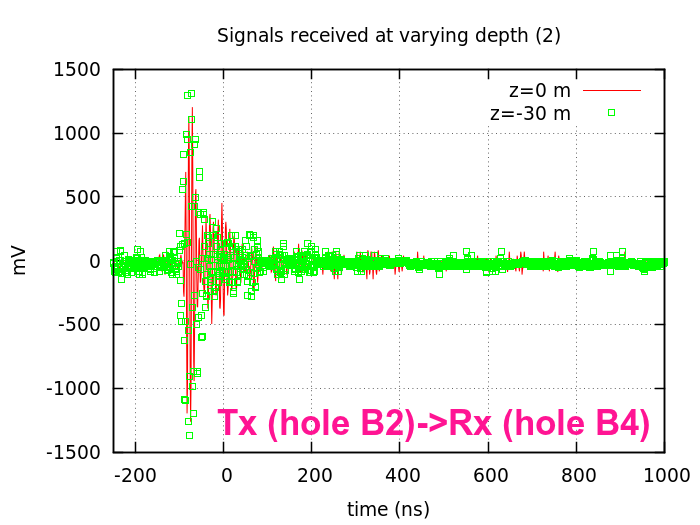}\caption{\it RICE hole B2 Tx$\to$RICE hole B4 waveform comparison.}\label{fig:B2_X_B4}
\end{minipage}
\end{figure}
Two different trials were carried out in the original experiment, conducted in 2003, which we have recently re-analyzed. That experiment targeted a measurement of the index-of-refraction dependence with depth, rather than measurement of surface waves, so control of systematics were designed with the former goal in mind. In each trial, two antennas were placed near-surface, and the propagation time, as well as the signal voltage registered at the receiver recorded. Next, the two antennas were co-lowered into parallel iceholes with a horizontal separation of order 50 m. Of interest here is the signal amplitude after the antennas have descended several wavelengths into the ice, at which point the received signal voltage can be compared with the near-surface measurements. If signal has been trapped in the boundary layer in the latter case, then, given the negligible radio-frequency signal absorption in the near-surface ice firn, the ratio of the near-surface signal amplitude relative to the in-ice signal amplitude should vary as $\sqrt{1/r}/\sqrt{1/r^2}$, or simply $\sqrt{r}$. I.e., the near-surface received amplitudes should be markedly larger than the sub-surface received amplitudes. Ralston's calculations\cite{Ralston}, in fact, predict an additional enhancement by a factor of $2\sqrt{2}$. The results of this exercise are shown in Figures \ref{fig:rodwell_X_B2} and \ref{fig:B2_X_B4}, for broadcasts between a) the South Pole Station (SPS) rodwell hole (evacuated volume remaining after melted ice has been extracted for general SPS use) and RICE hole B2 and b) RICE hole B2 and RICE hole B4, respectively. Although there is some moderate change in the waveform shapes of the near-surface vs. the deep-ice received signals, the overall received power 
in the first $\sim$100 ns of the initial pulse emission is comparable in the in-ice vs. the near-surface transmission cases, and inconsistent with the expected enhancement by a factor of $>\sqrt{50~m}$ that would be expected for the case of surface signal flux trapping.

\subsection{Check of Surface Confinement using a Snow `Barrier'} 
Although not fully examined in the original theoretical work, it is interesting to investigate the possibility that flux may be confined along a contoured surface. At South Pole, an additional check was performed wherein the signal received between two horn transmitters, separated by 675 m and facing each other on the snow surface was compared. In the first configuration, the two horn antennas had clear line-of-sight (LOS) relative to each other; in that configuration, the receiver antenna registered signals at Signal-to-Noise Ratio (SNR) levels, in units of the rms noise voltage $\sigma_{thermal}$, of 12$\sigma_{thermal}$ and 14$\sigma_{thermal}$ for the HPol and VPol configurations, respectively. In the second configuration, although the separation distance remained at 675 m, the transmitter was displaced horizontally approximately 100 meters, to a location where the snow surface was approximately 4 meters lower compared to the original location, such that there was no longer LOS to the receiver. In that second case, we observe no signal above noise, ruling out the possibility that a surface wave is following the contour of the surface between the transmitter source and the receiver.

\subsection{Amplitude Measurements at South Pole with the ARA Experiment}
A successor to RICE, the South Polar Askaryan Radio Array\cite{ARA} (ARA) neutrino detection experiment similarly consists of an array of radio receivers deployed both on the surface, as well as in several boreholes, separated by $\sim$20 m horizontally, with each hole containing 3--4 vertically stacked receiver antennas, separated by 5--10 m in the bulk ice. ARA has, thus far, developed in three stages: a) deployment of a prototype (`testbed') in January, 2011, consisting of 2 surface antennas sensitive over the frequency range $f<200$ MHz, 4 near-surface antennas deployed in the upper meter of Antarctic ice sensitive over 150 MHz$<f<$900 MHz, plus 10 antennas (150 MHz$<f<$900 MHz) deployed between 30 and 24 meters beneath the surface, b) deployment of the ARA1 station in January, 2012, including 4 surface antennas, and using optical fiber to convey signals from the 16 $\sim$90-m deep in-ice antennas to the surface, and c) deployment of the ARA2 and ARA3 stations in January, 2013, based on the ARA1 model, but with a deeper in-ice antenna deployment, to a depth of 180--200 meters. 

Of these, the testbed data are most informative for surface wave studies. Calibration continuous wave (CW) data were taken in the testbed receiver array in December, 2011, from two different transmitter configurations: a) RICE dipole transmitter buried just sub-surface (z=--25 cm) at a radial distance 100 m from the center of the testbed array, and b) the same dipole transmitter elevated 1 meter above the surface. For both of these cases, ARA receiver data were taken at 90-degree angular increments, in order to average over possible azimuthal Tx or Rx asymmetries. 

\begin{figure}[ht]
\begin{minipage}{3.1in}
\includegraphics[width=\textwidth,angle=0]{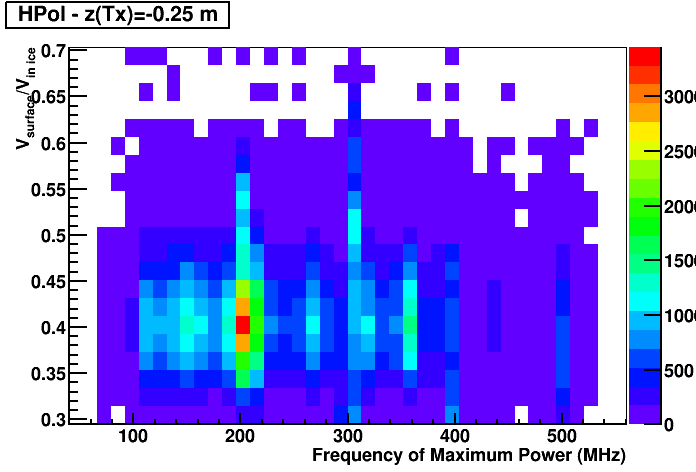}\caption{\it Ratio of total signal voltage in two HPol testbed surface antennas ($-1<z<0$m) to six HPol in-ice antennas ($-30<z<-22$ m) as a function of frequency; dipole transmitter located 25 cm below ice surface and broadcasting in 100 MHz increments.} \label{fig:TB_Tx_sub_HPol}
\end{minipage}
\hspace{0.3in}
\begin{minipage}{3.1in}
\includegraphics[width=\textwidth,angle=0]{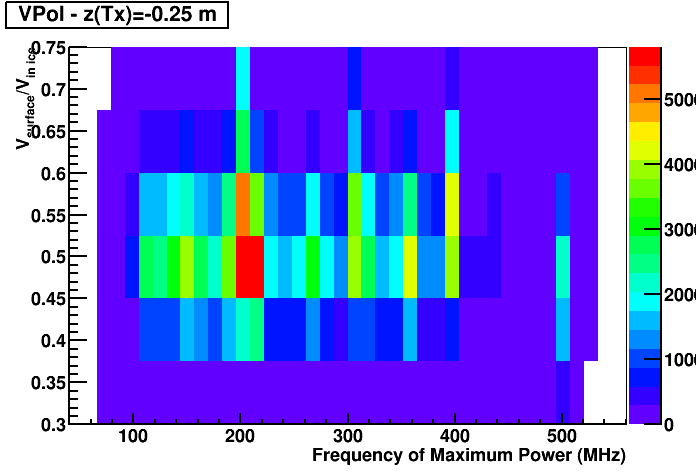}\caption{\it Ratio of total signal voltage in two VPol testbed surface antennas ($-1<z<0$m) to four VPol in-ice antennas ($-30<z<-22$ m) as a function of frequency; dipole transmitter located 25 cm below ice surface and broadcasting in 100 MHz increments.}\label{fig:TB_Tx_sub_VPol}
\end{minipage}
\end{figure}

For configuration a), both HPol and VPol data are shown, as a function of transmitter frequency, broadcast in 100 MHz increments, in Figures \ref{fig:TB_Tx_sub_HPol} and \ref{fig:TB_Tx_sub_VPol}. Shown is the ratio of peak signal amplitude in the frequency domain (after performing a Fourier transform on the time-domain captured waveforms), averaged over the 4 compass-point Tx source locations for which data were taken. To examine possible surface wave effects, and to remove possible transmitter systematics, we track the ratio of the response in the surface receiver antennas to the response in the in-ice antennas.

\begin{figure}[ht]
\begin{minipage}{3.1in}
\includegraphics[width=\textwidth,angle=0]{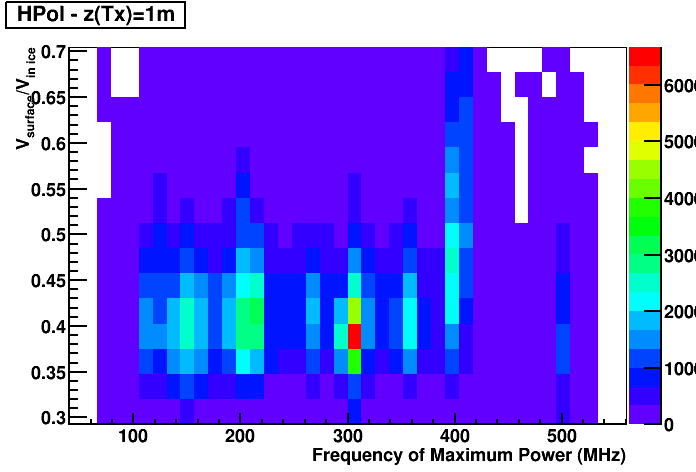}\caption{\it Ratio of total signal voltage in two HPol testbed surface antennas ($-1<z<0$m) to six HPol in-ice antennas ($-30<z<-22$ m) as a function of frequency; dipole transmitter located 1 m above ice surface and broadcasting in 100 MHz increments.} \label{fig:HPol_elevated_Tx}
\end{minipage}
\hspace{0.3in}
\begin{minipage}{3.1in}
\includegraphics[width=\textwidth,angle=0]{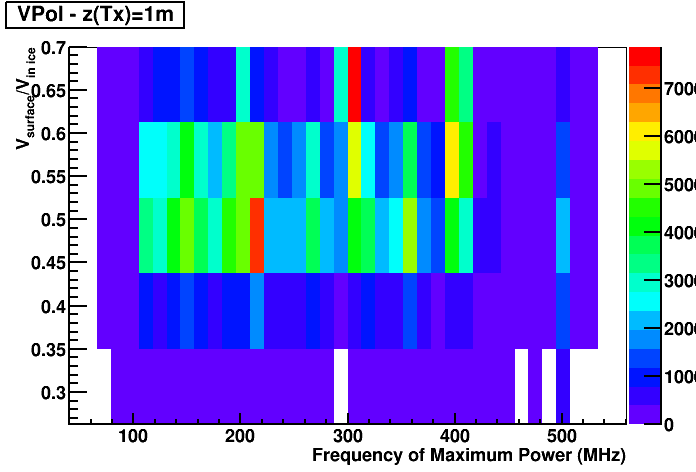}\caption{\it Ratio of total signal voltage in two VPol testbed surface antennas ($-1<z<0$m) to four VPol in-ice antennas ($-30<z<-22$ m) as a function of frequency; dipole transmitter located 1 m above ice surface and broadcasting in 100 MHz increments.}\label{fig:VPol_elevated_Tx}
\end{minipage}
\end{figure}

\begin{figure}[ht]
\begin{minipage}{3.1in}
\includegraphics[width=\textwidth,angle=0]{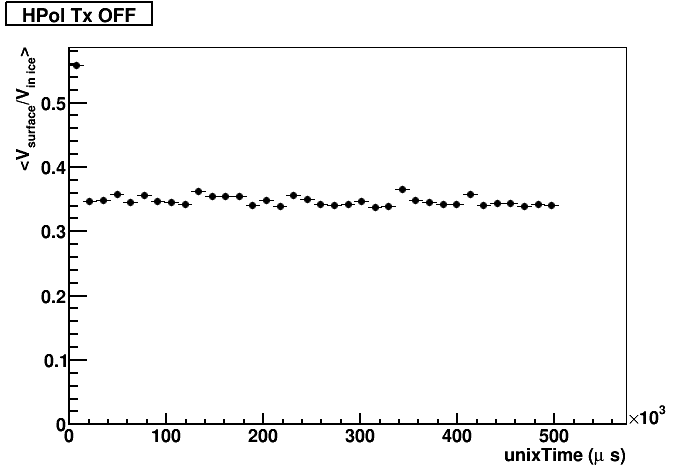}\caption{\it Ratio of total signal voltage in two HPol testbed surface antennas ($-1<z<0$m) to six HPol in-ice antennas ($-30<z<-22$ m) as a function of time within the GPS second; population near zero corresponds to reception of local calibration pulser firing on the GPS second.} \label{fig:HPol_TXOFF}
\end{minipage}
\hspace{0.3in}
\begin{minipage}{3.1in}
\includegraphics[width=\textwidth,angle=0]{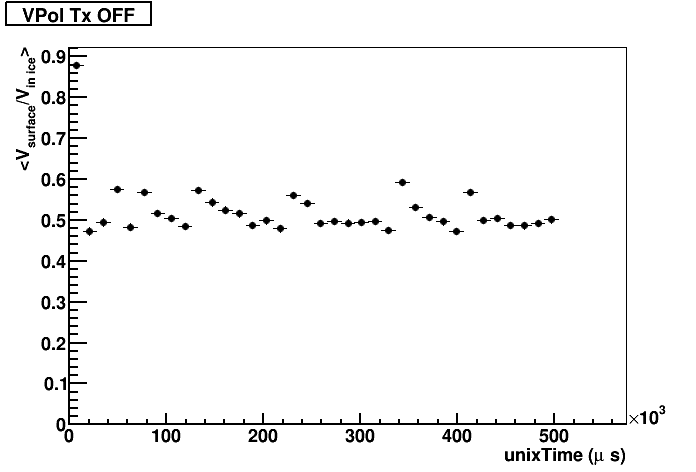}\caption{\it Ratio of total signal voltage in two VPol testbed surface antennas ($-1<z<0$m) to four VPol in-ice antennas ($-30<z<-22$ m) as a function of time within the GPS second; population near zero corresponds to reception of local calibration pulser firing on the GPS second.}\label{fig:VPol_TXOFF}
\end{minipage}
\end{figure}

We expect evidence for surface confinement of radio-frequency signal in Figs. \ref{fig:TB_Tx_sub_HPol}, \ref{fig:TB_Tx_sub_VPol}, \ref{fig:HPol_elevated_Tx} and \ref{fig:VPol_elevated_Tx}  to manifest itself as an enhanced voltage ratio obtained when broadcasting from a near-surface transmitter location, when compared with the voltage ratio obtained when the transmitter is off, and based on `thermal' noise trigger events (Figs. \ref{fig:HPol_TXOFF} and \ref{fig:VPol_TXOFF}). In fact, all data show roughly comparable ratios for all measured configurations, inconsistent with any surface wave coupling.

\subsection{Sand Measurements made at Great Sand Dunes National Park (CO)}
The large-area sand deposits comprising the Great Sand Dunes National Park (GSDNP) in Colorado provide a nearly semi-infinite surface dielectric, which, in principle, allows discrimination between $1/r$ and $1/\sqrt{r}$ amplitude spreading. Field measurements were made within the Park, at Latitude 37$^\circ$43'58''N and Longitude 105$^\circ$30'44''W, using a pair of GHz-band horn antennas designed specifically for this measurement. A large, flat dune, roughly 30 meters high was selected for the measurement, with a smooth surface of uniform sand tens of meters deep.

\begin{figure}[h]
\centering
\includegraphics[width=.8\textwidth]{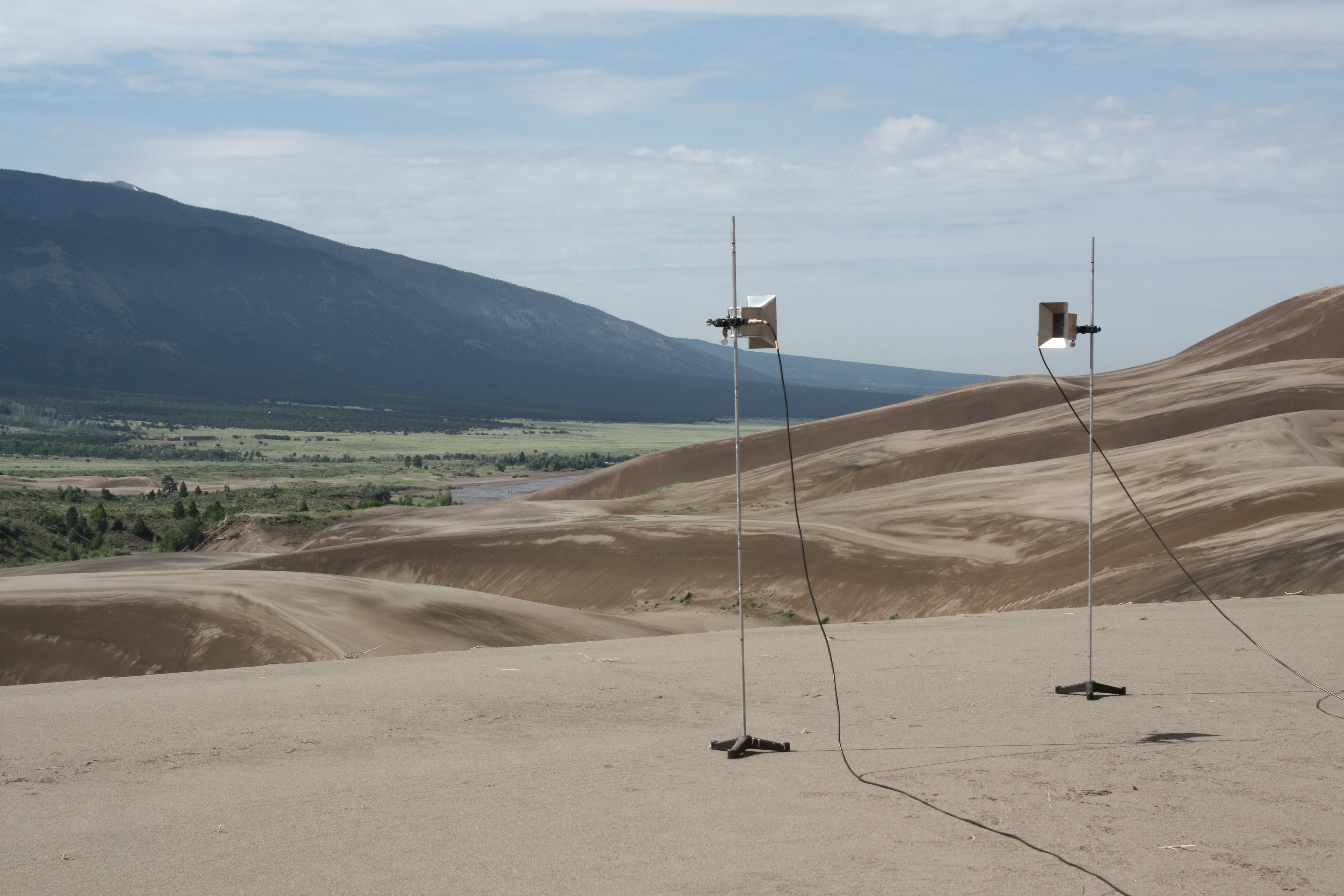}
\caption{\it Photograph of the two horn antennas used at the Great Sand Dunes National Park in their ``free-space'' configuration.\label{antennas}}
\label{fig:antennas}
\end{figure}

\begin{figure}[h]
\centering
\includegraphics[width=.6\textwidth]{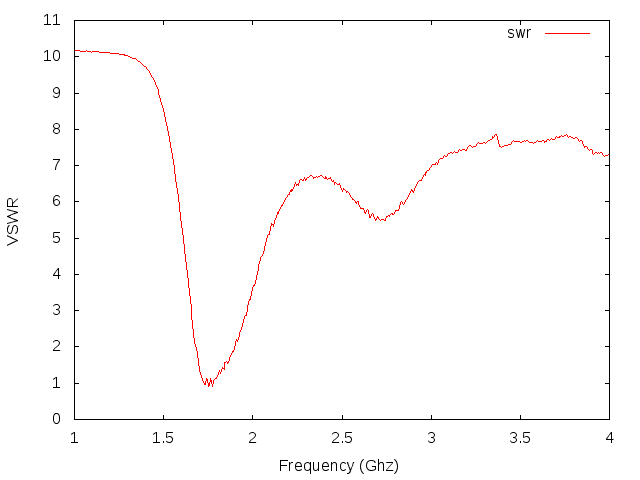}
\caption{\it Measured Voltage-Standing Wave Ratio (VSWR) for horn antennas used in the GSDNP experiment.\label{swr}}
\end{figure}

The two antennas visible in Figure \ref{fig:antennas} are aluminum, GHz-band antennas, with voltage standing wave ratio (VSWR) as shown in Figure \ref{swr}. For this test they were mounted on metal stands and connected, via 16 meters of coaxial cable, to a network analyzer. The two mounted antennas were initially placed a distance $d$ of 2.2 meters apart, ensuring that the receiver was in the far field for the entire band of 1--4 GHz. Measurements of received power were then taken at 19 discrete separation distances corresponding to this initial position plus $d/2$, and were repeated for different antenna configurations. 
Note that these data were somewhat compromised by the presence of moisture at depths tens of cm into the 
sand, however, our studies indicate relative insensitivity to this moisture layer, based on data taken after an extended
period of dry weather vs. data taken directly after a rain storm.

\begin{figure}[h]
\centering
\includegraphics[width=.8\textwidth]{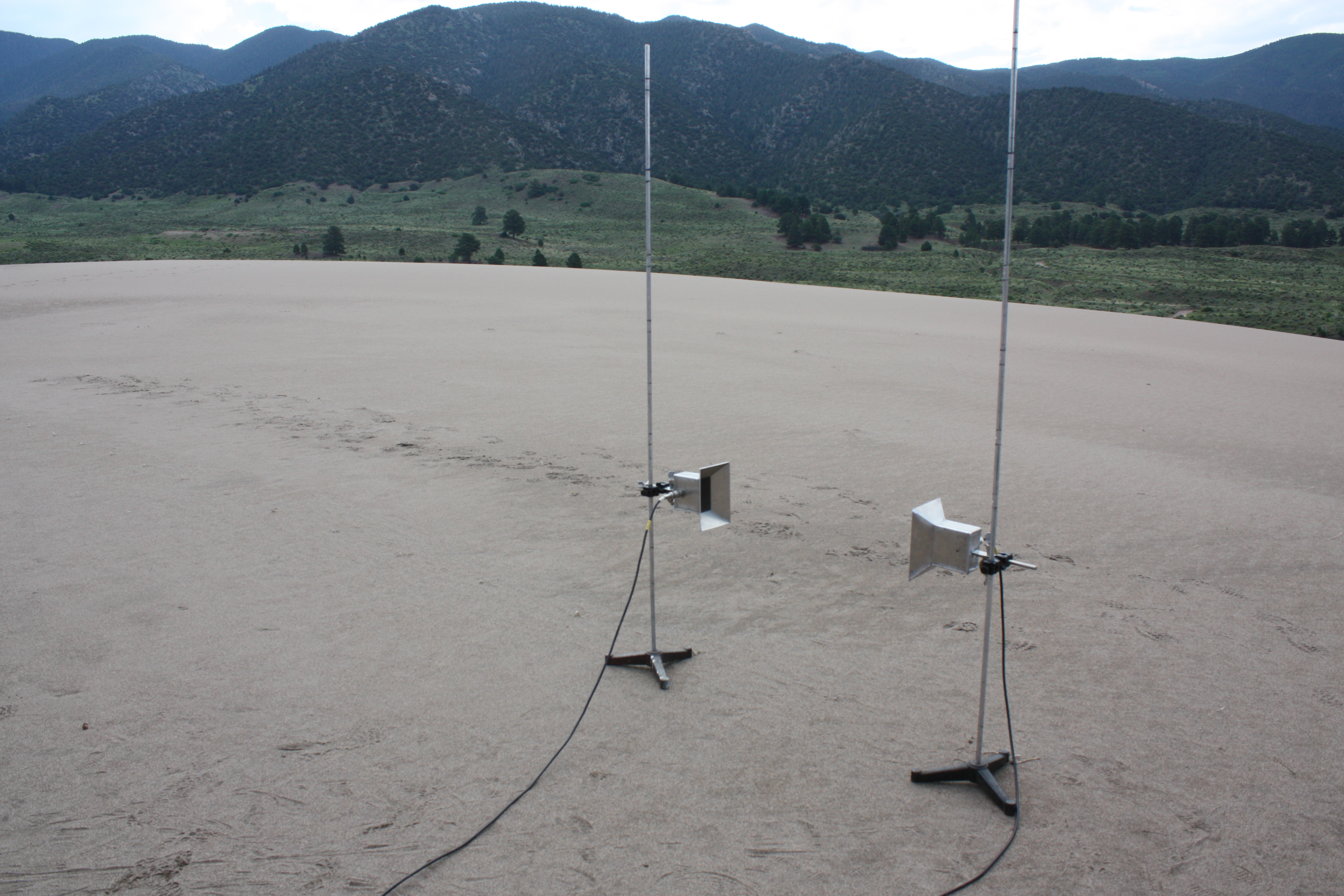}
\caption{\it The two horn antennas used for the Great Sand Dunes National Park experiment, in a near-surface setup.\label{fig:antennas2}}
\end{figure}

Three trials were compared. In each trial, the network analyzer (Anritsu-Wiltron MS2304A) was used in S12 mode, with our identical, custom-made horn antennas connected to the transmitter and receiver ports of the network analyzer, respectively. The signal strength was measured, as a function of separation between transmitter and receiver for the case where: a) both transmitter and receiver are in-air, at least one wavelength above the sand-air surface, b) transmitter is buried in the sand, and the receiver is in the air, at least one wavelength above/below the sand-air interface ($\lambda_{air}$=7.5 cm at 4 GHz), and c) both transmitter and receiver are each half-submerged in the sand, which has a complex refractive index at microwave frequencies of 2.288 -- 0.057i, giving $\lambda_{sand}$ approximately 3 cm at 4 GHz and also implying negligible signal absorption at our typical separation distances. Cases a) and b) should, for a properly calibrated system, exhibit $1/r$ flux spreading. If waves are confined to the surface, we expect $1/\sqrt{r}$ spreading for case c). Electrically invisible position indicators (very small pieces of light wood just at the surface of the sand) were placed out 20 cm out of the specular beam path, as shown in Figure \ref{fig:antennas2}.

\subsubsection{Free Space}
\begin{figure}[h]
\centering
\includegraphics[width=.6\textwidth]{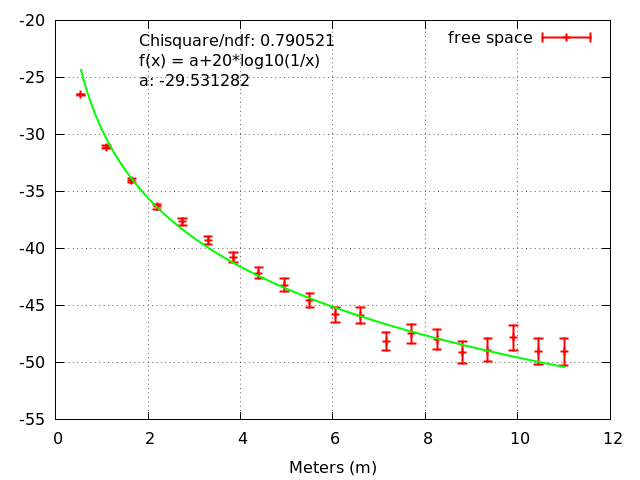}
\caption{\it Fit to the free-space data with overlay to $1/r$ functional form.\label{fig:freespace_results}}

\end{figure}

The first measurement taken was a free space test, to verify the expected flux spreading in three-dimensions. The antennas were placed at a height of 2 m, as shown in Figure \ref{fig:antennas}, greatly reducing the influence of the sand surface on the measurement. The results of this test for several frequency bins are shown in Figure \ref{fig:freespace_results}, with a fit to $1/r$ overlaid. We observe excellent agreement with the $1/r$ hypothesis.

\subsubsection{Surface Measurements}
\begin{figure}[h]
\centering
\includegraphics[width=.6\textwidth]{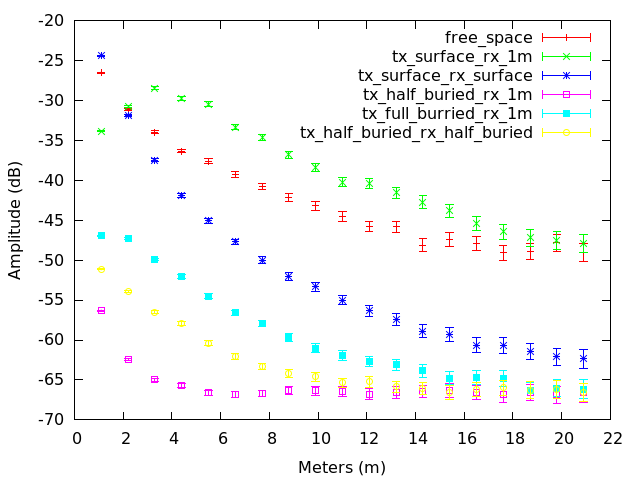}
\caption{\it A comparison of several measurements. Note that the noise floor of the Network Analyzer is approximately --65 dB; there are therefore non-negligible contributions to the observed received signal up to $\sim -$60 dB\label{fig:all_msrmnts}}
\end{figure}

\begin{figure}[h]
\centering
\includegraphics[width=.6\textwidth]{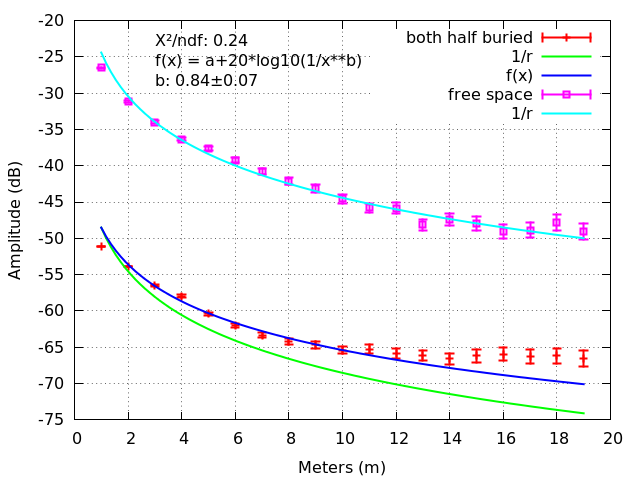}
\caption{\it Great Sand Dunes National Park test of surface wave coupling: comparison of free space versus tx buried, rx 1m in height. \label{fig:compare}}
\end{figure}

The remainder of the measurements were taken to probe the possibility of exciting surface waves. The transmitter and receiver were set at different heights, including i) sitting on the surface, and in contact with the surface, ii) half-buried, iii) fully-buried, or iv) elevated at least one wavelength above the surface. A complete compilation of our obtained data are shown in Figure \ref{fig:all_msrmnts}; a direct comparison of the half-buried data with free space are shown in \ref{fig:compare}.
Overall, there was no significant observed difference, besides a reduction in overall amplitude (which we attribute to the Fresnel transmission coefficient through the air-sand interface) between the free space case and the case where the transmitter was fully buried and the receiver was placed 1 m off the surface.
We can fit our measured amplitude points to the form $1/r^b$.
 The case of both transmitter and receiver half-buried shows an amplitude falloff which favors $1/r$ over 1/$\sqrt{r}$, although it should be noted that this measurement is close to the noise floor, particularly for distances greater than 8 m. separations; we have correspondingly restricted the fit to the first 8 points in Figure \ref{fig:compare}. The exponent obtained in Fig. \ref{fig:compare}, for example, increases towards 0.9 as we reduce the number of fitted points, albeit with larger error bars.

\section{Studies of the real part of the permittivity}
As discussed previously, it has been suggested that surface waves may travel with a velocity (both $v_g$ and also $v_p$) corresponding
to $n<1$. We have therefore searched for apparently superluminal signal velocities. 
To directly measure the index-of-refraction at dielectric boundaries, a radio-frequency continuous wave (CW) signal was generated using a Rhode and Schwarz SMHU signal generator (0.1 to 4320 MHz) with the 19 dBmW output connected to an 8 cm, 14 gauge copper half-wavelength dipole antenna, via a 2 m (CableWave) N-connectorized coaxial cable soldered to the antenna feed point.  By measuring signals from a stationary and moving receiver, the phase at distances corresponding to half-integer multiples of the observed wavelength ($\lambda$) were recorded, yielding the effective phase velocity index-of-refraction ($n = c/f\lambda$).  The receiving dipoles were soldered to 3 m coaxial cables and connected to a Tektronix TDS 5104 oscilloscope.  In addition to varying the distance between the two receivers, the height of both receivers above the surface was varied.  As outlined above, distances between the two receivers corresponding to both $0^{\circ}$ and $180^{\circ}$ phase shifts were recorded.

\begin{figure}
\begin{centering}
	\includegraphics[width=0.8\textwidth]{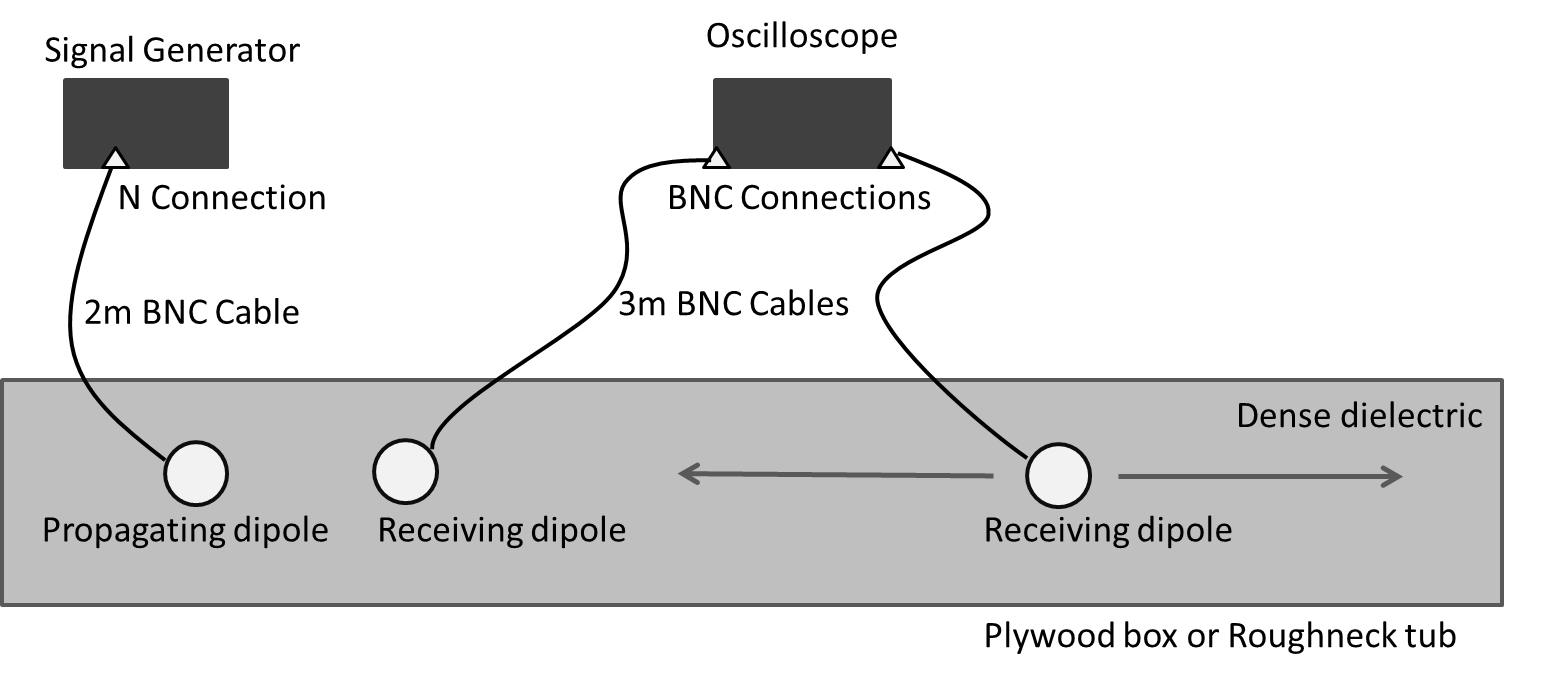}
	\caption{\it Geometry for measurement of surface wave phase velocities.  A signal propagating from a dipole creates an electromagnetic wave with constant frequency; the wavelength is measured by moving one receiver with respect to both a static receiver and the transmitter. \label{basicSetup}}
\end{centering}
\end{figure}

This procedure was repeated several times over the length of the dielectric medium to minimize statistical fluctuations.  The two-dimensional surface materials tested were water, water doped with pool salt to varying salinities, granulated fused silica (sand), and granulated sodium chloride (salt).  CW signal frequencies between 750 and 1500 MHz were investigated, each frequency being within the efficient transmission region of the antennas.  A box to contain the dielectric media (except for the doped water) was constructed with 8.9 cm x 61 cm x 122 cm dimensions.  Sand measurements from 150--500 MHz were also performed in an outdoor sand pit (c.f. Fig. \ref{vswr}) with the lower-frequency RICE dipole antennas\cite{RICE}.  The purified water was contained within a circular pool roughly 3 m in diameter, and the salinity was recorded on an ExTech EC400 conductivity/TDS/salinity meter.  The technique is summarized in Figure \ref{basicSetup}.

\subsection{Laboratory Wave Speed Measurements with Sand and Salt}
\begin{figure}
\begin{centering}
	\includegraphics[width=0.8\textwidth]{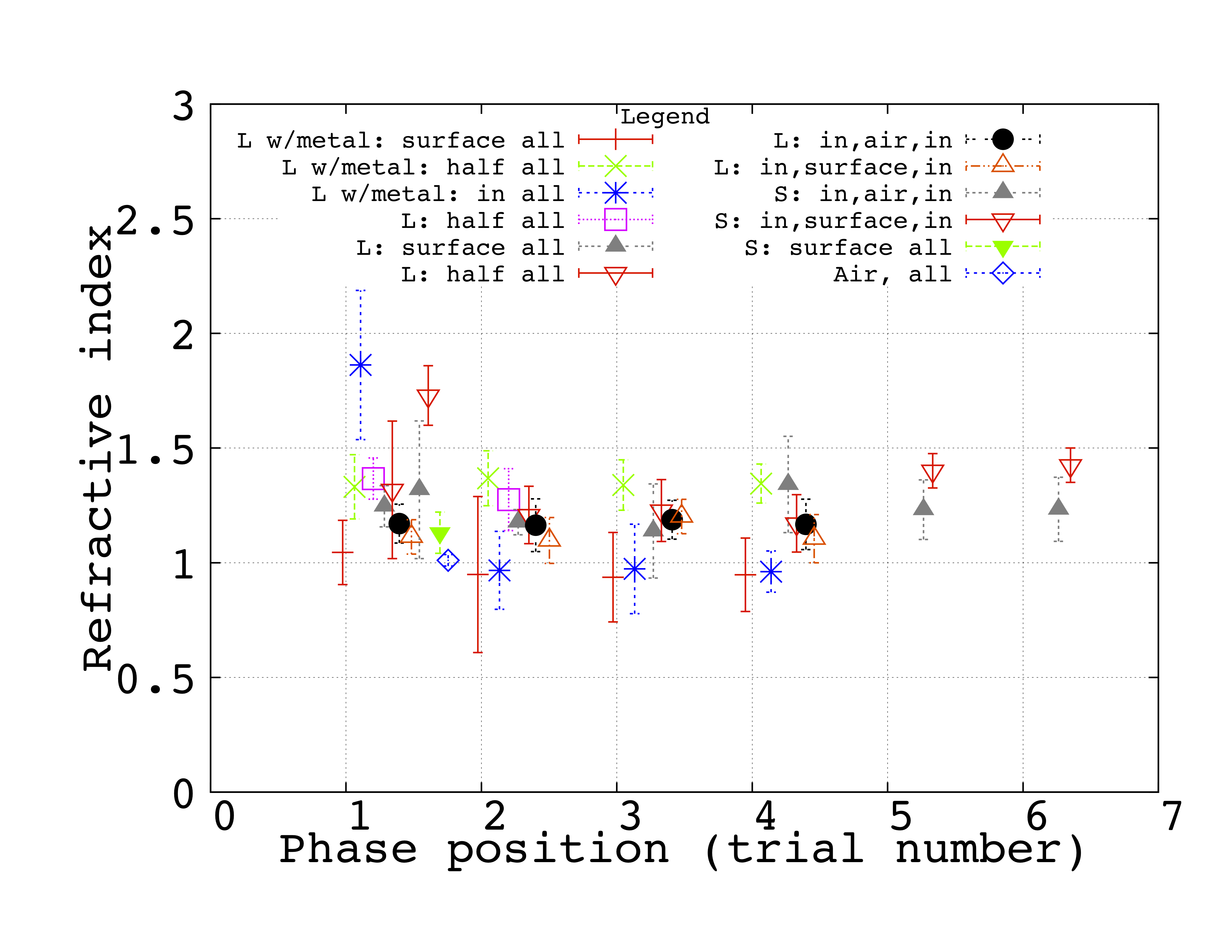}
	\caption{\it Measured refractive indices of various propagation paths near the sand surface at 1000 MHz, along with the same result for air. The x-axis represents the phase position of the moveable receiver, and the y-axis is the measured refractive index. The points have been spread out in the x-direction for visual clarity; each point corresponds to the phase position to the immediate left. The label \textbf{L} indicates that the index measurements were taken in the standard experimental box, while \textbf{S} indicates an earlier, smaller container.  The smaller container was investigated initially, and found to admit only 1-2 phase positions. \label{fig1}}
\end{centering}
\end{figure}

\begin{figure}
\begin{center}
	\includegraphics[width=0.8\textwidth]{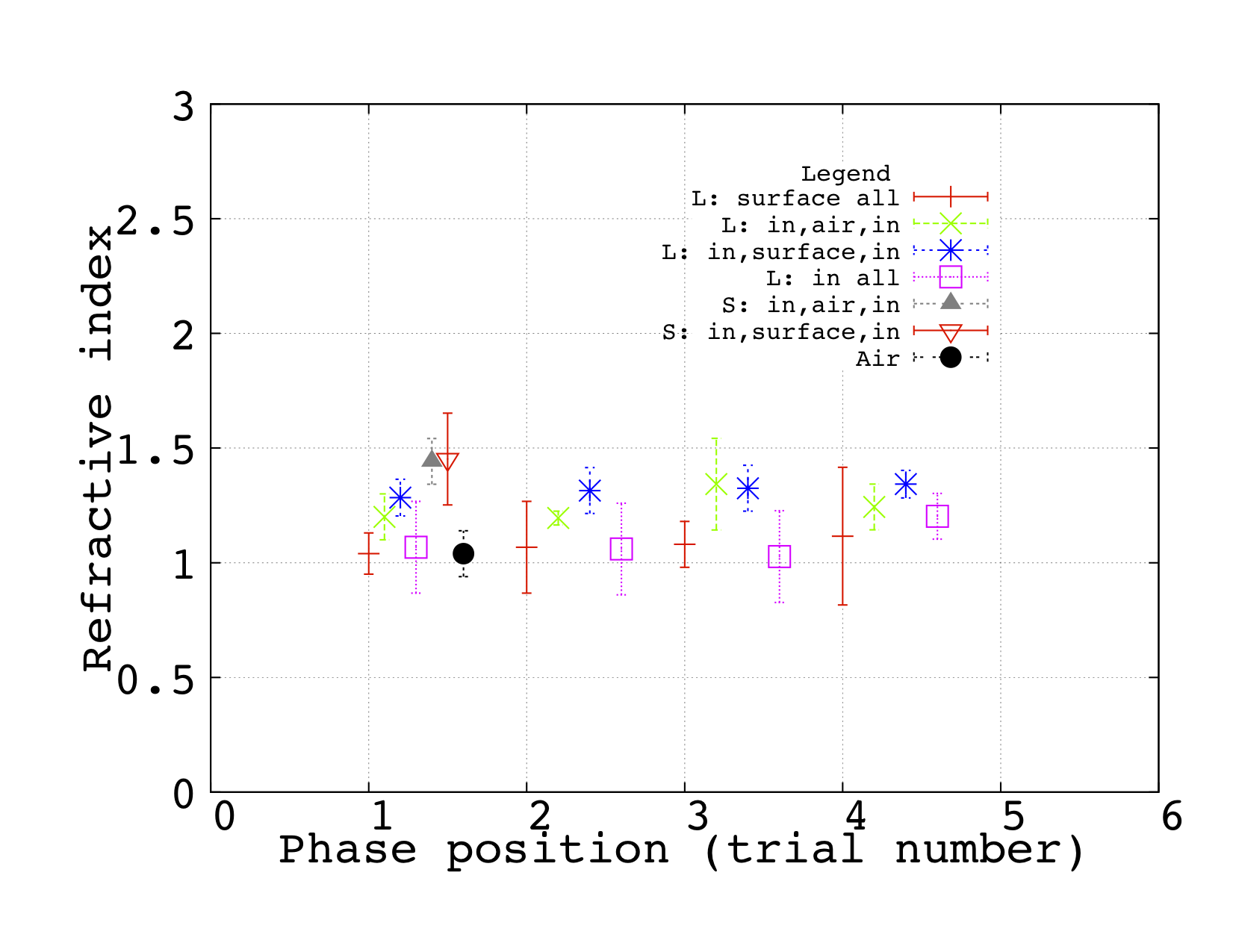}
	\caption{\it The measured refractive indices for various propagation paths through sand at 1500 MHz, along with the same result for air. The x-axis represents the phase position of the dynamic receiver, and the y-axis is the measured refractive index. The points have been spread out in the x-direction for visual clarity; as before, each point corresponds to the phase position to the immediate left.  \label{fig2}}
\end{center}
\end{figure}

\begin{figure}
\begin{center}
	\includegraphics[width=0.8\textwidth]{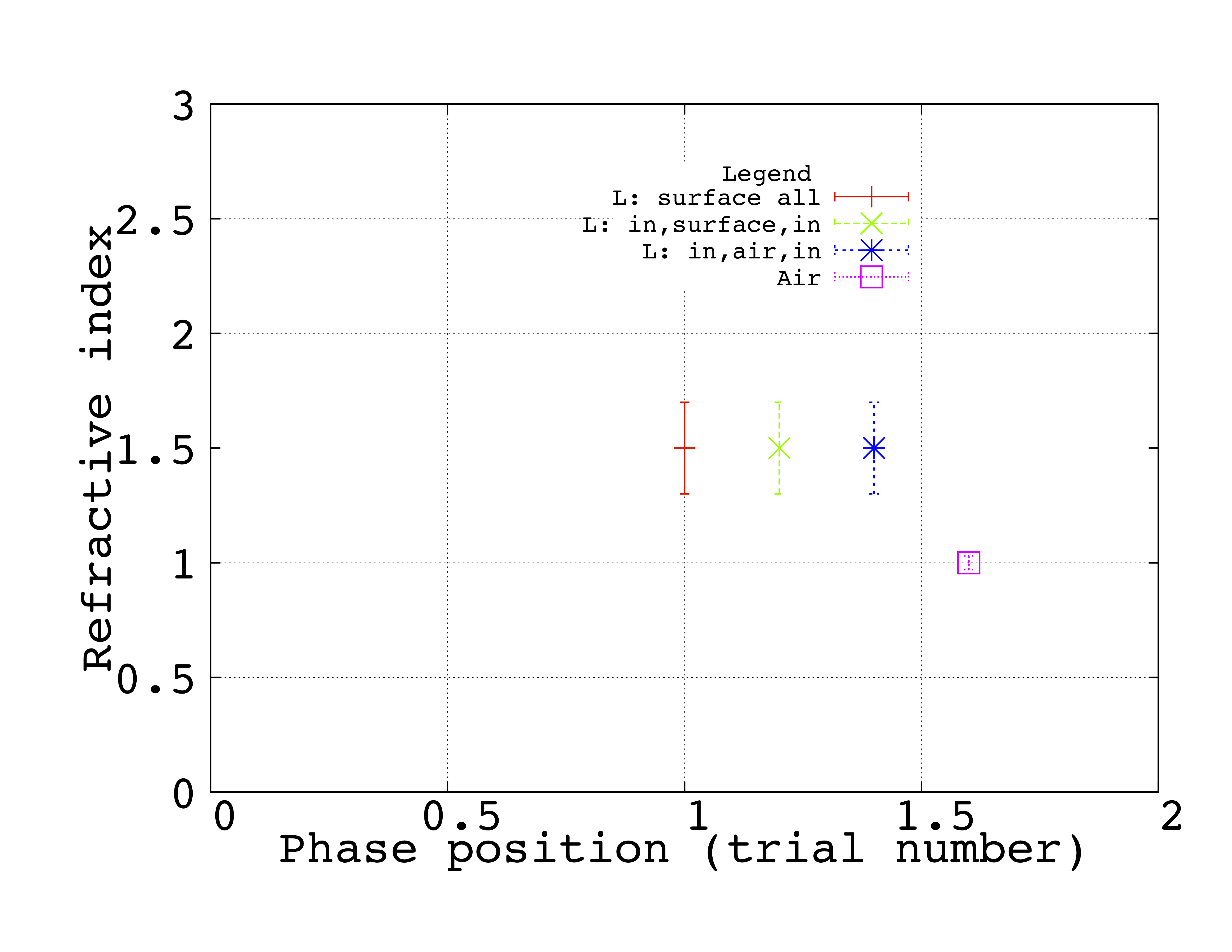}
	\caption{\it The refractive indices for various propagation paths through salt at 1000 MHz.  Only the first phase position was tested (x-axis); the y-axis is the measured refractive index.  A free-space measurement was performed as a cross-check (square point), yielding a value consistent with 1, as expected. \label{fig3}}
\end{center}
\end{figure}

\begin{figure}
\begin{center}
	\includegraphics[trim=1cm 1.8cm 1cm 2.1cm,clip=true,width=10cm]{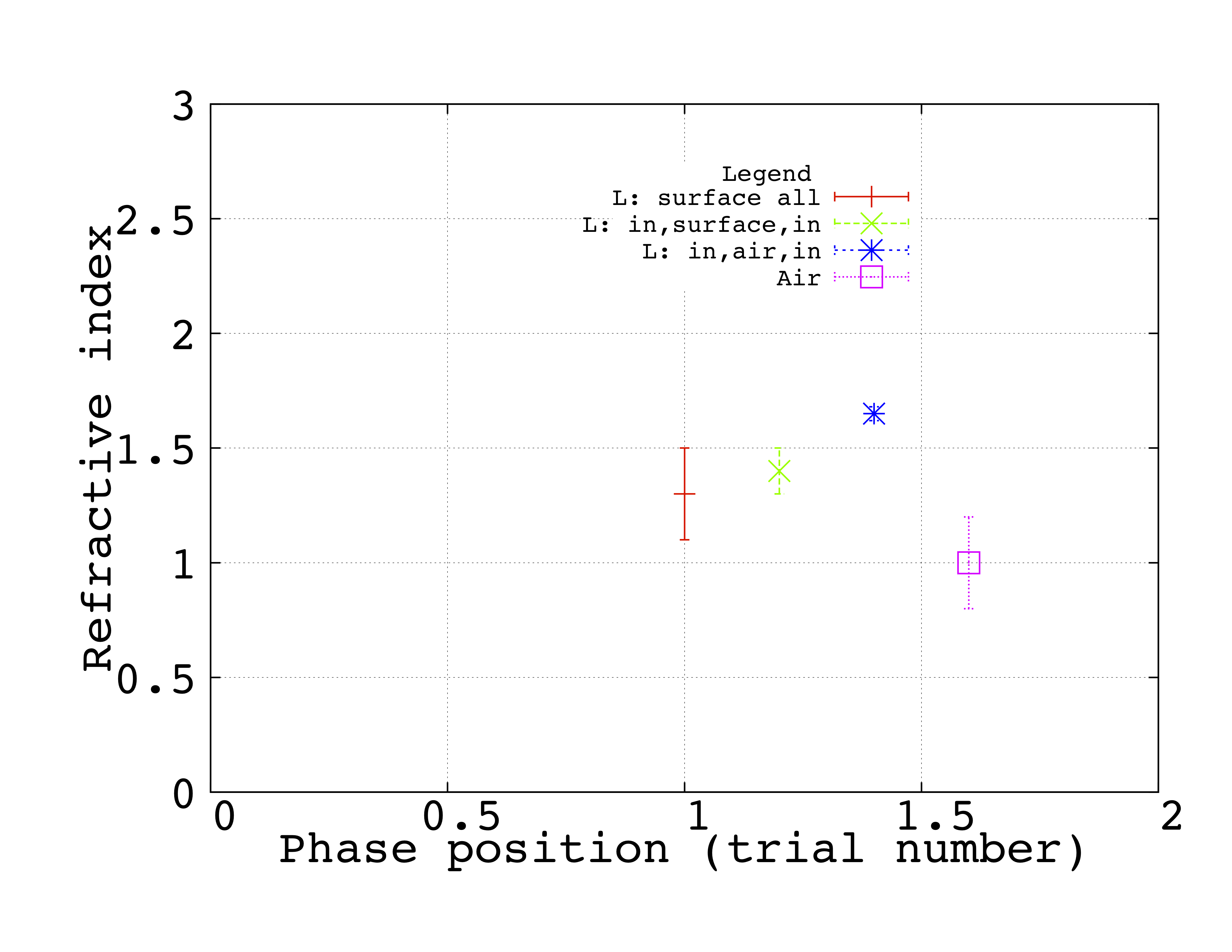}
	\caption{\it The measured refractive indices of various propagation paths through salt at 1500 MHz.  Only the first phase position was tested (x-axis); y-axis is the measured index-of-refraction.  A free-space measurement was performed as a cross-check (the square point). Errors shown are statistical only.\label{fig4}}
	\end{center}
\end{figure}

Antenna orientations were sorted into the categories of \textbf{in}, \textbf{surface}, \textbf{air}, and \textbf{half}. The \textbf{in} label indicates antenna placement fully within the dense dielectric, \textbf{half} indicates halfway within the dense dielectric, \textbf{surface} indicates placement directly on top of the dense dielectric, and \textbf{air} indicates antenna placement one antenna-length above the dense dielectric. A single antenna placement descriptor indicates all antennas were in the same position with respect to the surface, and ordered placement descriptors describe antenna position in propagation order (transmitter, static receiver, and moveable receiver, respectively). The label \textbf{w/metal} indicates sand doped with a 10:1 ratio of sand to aluminum shavings. In Figures \ref{fig1}-\ref{fig4}, the x-axis indicates the measurement trial number, corresponding to the phase position (in units of wavelengths) of the moving receiver.  As that receiver was translated farther from the static receiver, the distance to each phase position was recorded, and the phase velocity index-of-refraction derived.  

Figure \ref{fig1} displays the refractive index results at a frequency of 1000 MHz; results are also summarized in Table \ref{tab1}.
All combinations of signal path are shown, with mean values from multiple trials presented along with one standard deviation statistical errors.  For laboratory measurements, Quikrete commercial sand (No. 1962) was used. Although the sand used in the outdoor sand pit (described below) was not commercial, the density is observed to be roughly the same.  Notice from Fig. \ref{fig1} that measurements having differing antenna orientations, but for the same phase position are statistically consistent. Measurements were made out to the highest possible phase position before noise on the oscilloscope began to dominate the signals.  Figure \ref{fig1} contains both data from the initial smaller set up ($\textbf{S}$) and the larger one ($\textbf{L}$).  The average index-of-refraction for all measurements is $1.22\pm0.03$.  Considering only measurements with aluminum-doped sand, the average is $1.17\pm0.08$.  Excluding from this subset the $\textbf{half}$ position, the modified average is $1.1\pm0.1$.  Excluding the metal-doped measurements from the global average yields $<n>=1.24\pm0.03$.  
The index-of-refraction of dry sand is $\approx$ 1.6\cite{dry},
\message{http://geotest.tamu.edu/userfiles/245/Hong_microwave_sand.pdf}
 whereas air has an index of 1.0.  Thus, the data suggest that the waves sample the average of the two propagation media, consistent with the fact that one wavelength is of order or larger than the separation between an antenna and the dielectric surface.  Table \ref{tab1} summarizes the sub-categories of refractive index.  All measurements give index-of-refraction values intermediate between air and sand.

\begin{table}
	\begin{center}
	\begin{tabular}{c c}
		\hline
		Sub-category & average$\pm$ error in mean \\ \hline 
		All-measurements &  $1.22\pm0.03$ \\ \hline
		Metal-doped & $1.17\pm0.08$ \\ \hline
		Metal-doped (excl. $\textbf{half}$) & $1.1\pm 0.1$ \\ \hline
		Excl. metal-doped & $1.24\pm0.03$ \\ \hline
		Air & $1.01\pm0.03$ \\ \hline
	\end{tabular}
	\end{center}
	\caption{\it Summary of the average measured index-of-refraction for sand, from laboratory studies performed at 1000 MHz. Errors shown are statistical only.\label{tab1}}
\end{table}

Figure \ref{fig2} displays the refractive index results at a frequency of 1500 MHz, with the same labeling conventions as Fig. \ref{fig1}.  All tested combinations of signal path are shown, with measurements made out to the highest possible phase position.  Relative to the measurements performed at 1000 MHz, the results at 1500 MHz seem to indicate increased sensitivity to the orientation of the transmitters and receivers, consistent with the expectation that, for smaller wavelengths, we are increasingly probing smaller-scale non-uniformities.  Figure \ref{fig2} contains both data from the initial smaller set up ($\textbf{S}$) and also the larger one ($\textbf{L}$).  The data corresponding to submerged transmitters and receivers ($\textbf{In}$) produce an index closer to that of air, while an all-measurement average and an average of just the large box data are both consistent with the results at 1000 MHz.  Table \ref{tab2} summarizes the results in Fig. \ref{fig2}.

\begin{table}
	\begin{center}
	\begin{tabular}{c c}
		\hline
		Sub-category & average$\pm$error in mean \\ \hline
		All-measurements &  $1.20\pm0.03$ \\ \hline
		$\textbf{In}$, only & $1.09\pm0.03$ \\ \hline
		All (excl. $\textbf{In}$) & $1.28\pm 0.02$ \\ \hline
		Small box only & $1.447\pm0.005$ \\ \hline
		Large box only & $1.17\pm0.03$ \\ \hline
		Air & $1.0\pm0.1$ \\ \hline
	\end{tabular}
	\end{center}
	\caption{\it A summary of the average index-of-refraction measurements for sand, from the laboratory studies performed at 1500 MHz. Errors shown are statistical only. \label{tab2}}
\end{table}

The experimental procedure using a moveable and a static receiving antenna was repeated in the smaller version of the box, filled with Morton 100\% NaCl rock salt, with results shown in Figures \ref{fig3} and \ref{fig4}.  Several measurements of the index-of-refraction of natural rock salt yielded values between 2.4 and 2.8\cite{salt1,salt2} for the 100-1000 MHz regime, depending on frequency and the rock salt purity.  
For this experiment, the size of the salt container limited the analysis to only the first phase position.  Additionally, the rock salt crystals changed the outline of the relatively small dipole antennas, limiting the number of orientations tested. As expected, the average measured index-of-refraction is higher than for the case of sand, given the known larger bulk index-of-refraction for NaCl compared to SiO$_2$.

\begin{table}
	\begin{center}
	\begin{tabular}{c c}
		\hline
		Sub-category & average$\pm$error in mean \\ \hline
		All measurements &  $1.50\pm0.01$ \\ \hline
		Air & $1.0\pm0.03$ \\ \hline
	\end{tabular}
	\end{center}
	\caption{\it Summary of the average measured index-of-refraction for salt, from laboratory studies performed at 1000 MHz. Errors shown are statistical only. \label{tab3}}
\end{table}
	
\begin{table}
	\begin{center}
	\begin{tabular}{c c}
		\hline
		Sub-category & average$\pm$error in mean \\ \hline
		All measurements &  $1.45\pm0.01$ \\ \hline
		Air & $1.0\pm0.2$ \\ \hline
	\end{tabular}
	\end{center}
	\caption{\it Summary of the measured average index-of-refraction for salt, from laboratory studies performed at 1500 MHz. Errors shown are statistical only. \label{tab4}}
\end{table}

\subsection{Phase Measurements using Purified water with Varying Salinity}
\begin{figure}[ht]
\begin{minipage}{3.1in}
\begin{center}
	\includegraphics[width=7cm]{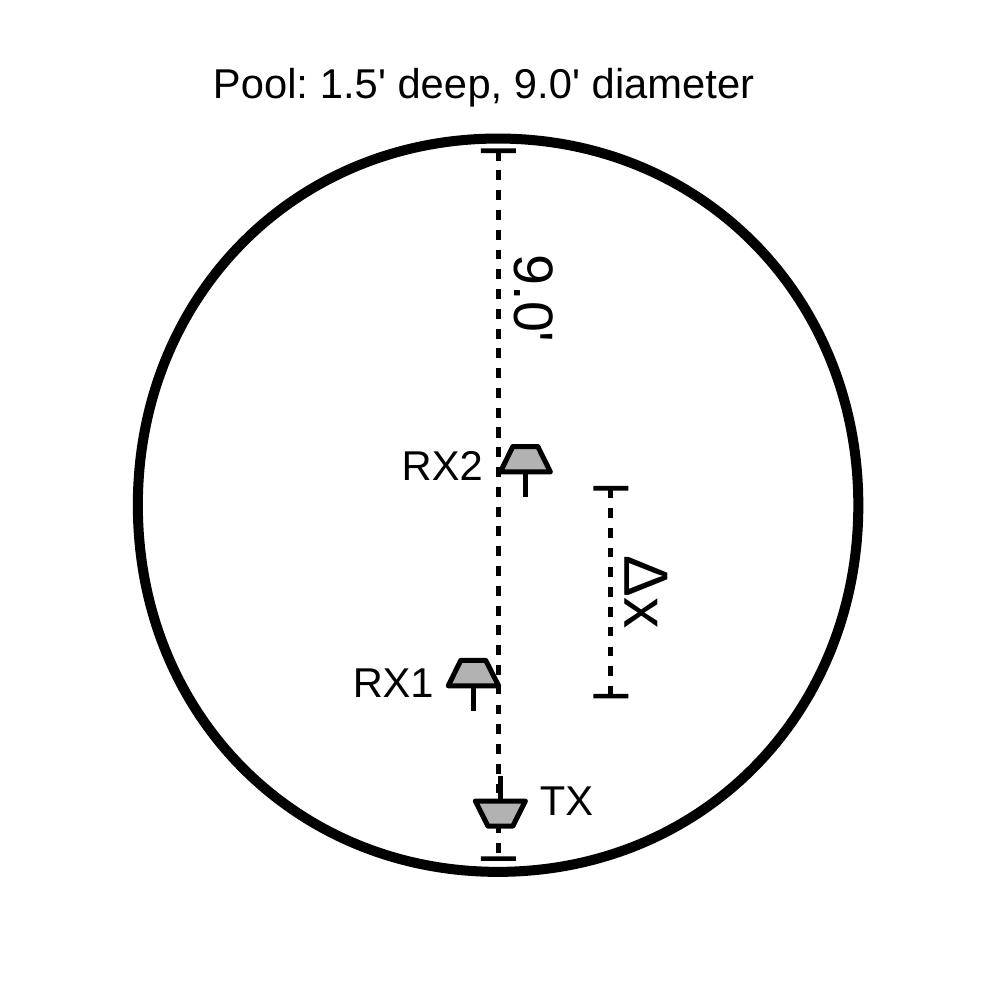}
	\caption{\it A diagram of the purified water setup.  The water was contained within a 3 m-diameter pool, with a water depth of 50 cm.  The transmitter antenna was located at one end, and a measuring tape was extended across the diameter to mark the position of the two receiving antennas.  \label{poolFig}}
	\end{center}
\end{minipage}
\hspace{0.3in}
\begin{minipage}{3.1in}
\begin{center}
	\includegraphics[width=8cm]{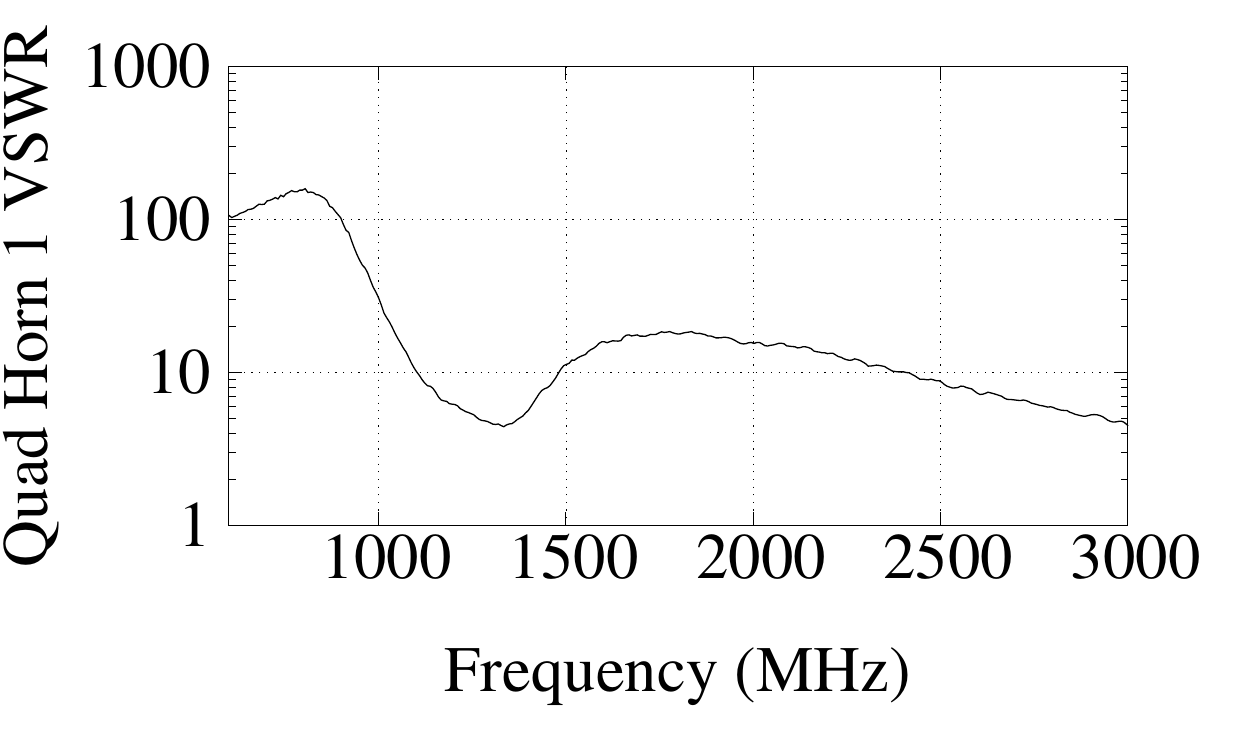}
	\caption{\it The VSWR of the small dipoles with quad horn attachments. \label{vswrs}}
	\end{center}
\end{minipage}
\end{figure}

The phased technique was also used to study the surface index-of-refraction for purified water doped with pool salt, at varying salinities.  It was quickly discovered that this setup (Fig. \ref{poolFig}) required higher gain antennas than the dipoles used previously, because of the higher intrinsic absorption of the medium.  Horn attachments were therefore employed to boost the gain of the simple dipole wires.  Figure \ref{vswrs} shows the VSWR versus frequency of the simple dipole wires with the quad attachment added.  It is important to note that the primary requirement for these CW measurements was that the relative phase be visible on the oscilloscope.  The antennas were placed above a 50 cm deep, 3 m diameter inflatable pool, filled with water purified via a charcoal filter through a hose attachment.  An ExTech EC400 conductivity/TDS/salinity meter showed that the purified water contained no measurable trace amount of salt before doping.

\begin{figure}
\begin{center}
	\includegraphics[width=10cm]{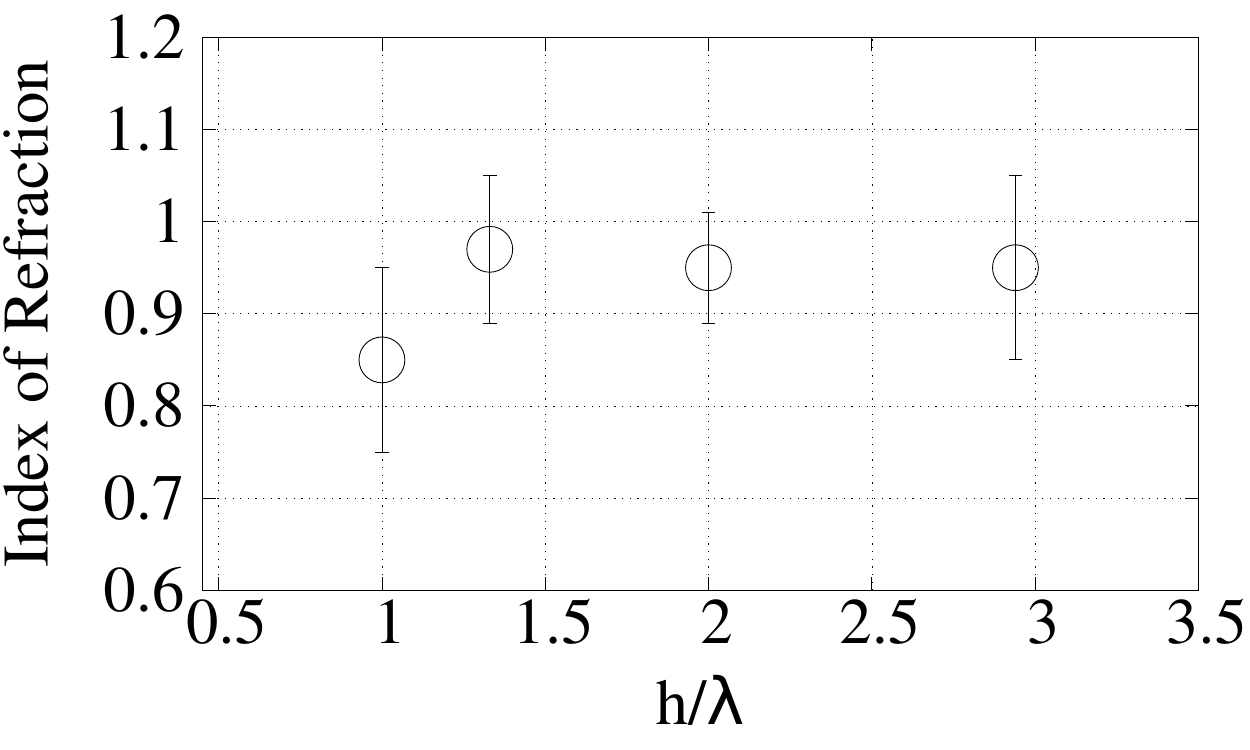}
	\caption{\it The measured surface index-of-refraction for water doped with salt ($7.16\pm0.04$ ppt), versus the ratio $h/\lambda$, where $h$ is the height above the surface, and $\lambda$ is the free space wavelength. \label{hlamb}}
	\end{center}
\end{figure}

The maximum range of the salinity meter was 10 parts per thousand (ppt) in added salt.  Typical sea water has $\approx 35$ g/kg in dissolved salts, or about 35 ppt.  The salinity of the pool was increased by adding pool salt incrementally, until the salinity reached 7.2 ppt.  The pool salt dissolved after stirring the pool, and no boiling was necessary to create a satisfactorily uniform solution.  The phased technique was then used to measure the index-of-refraction at varying heights above the surface.  Figure \ref{hlamb} shows the results at a high frequency (2000 MHz), corresponding to a suitably small (15 cm) wavelength.  Using frequencies between 1000-1500 MHz would have made the antennas more efficient, but limited the number of different height measurements due to the geometry of the setup.

Each of the data points represents three trials, each involving four phase positions (alternating betweeen 0$^{\circ}$ and 180$^{\circ}$).  The number of measurable phase positions was limited by our finite signal amplitude and non-zero scope noise (there was no amplifier in the system).  The errorbars in Fig. \ref{hlamb} show the standard deviations of the refractive index measurements, and the central values are the average across trials.  The letter $h$ designates the antenna height above the surface (44 cm, 30 cm, 20 cm, and 15 cm).  When the ratio $h/\lambda$ decreases to 1.0, the  index-of-refraction is statistically most distinct from $n=1.0$.  We note one point with an index-of-refraction apparently $2\sigma$ below 1.0, which, given our total number of trials, is not statistically significant, and, if real, a much smaller effect quantitatively than would have been predicted in \cite{Ralston}. For completeness, we point out that our result is also qualitatively not inconsistent with the claims of earlier experiments\cite{SU1,SU2,SU3}. 

\subsection{Outdoor Sand Pit Data}

\begin{figure}[htpb]
\begin{center}
	\includegraphics[width=8cm]{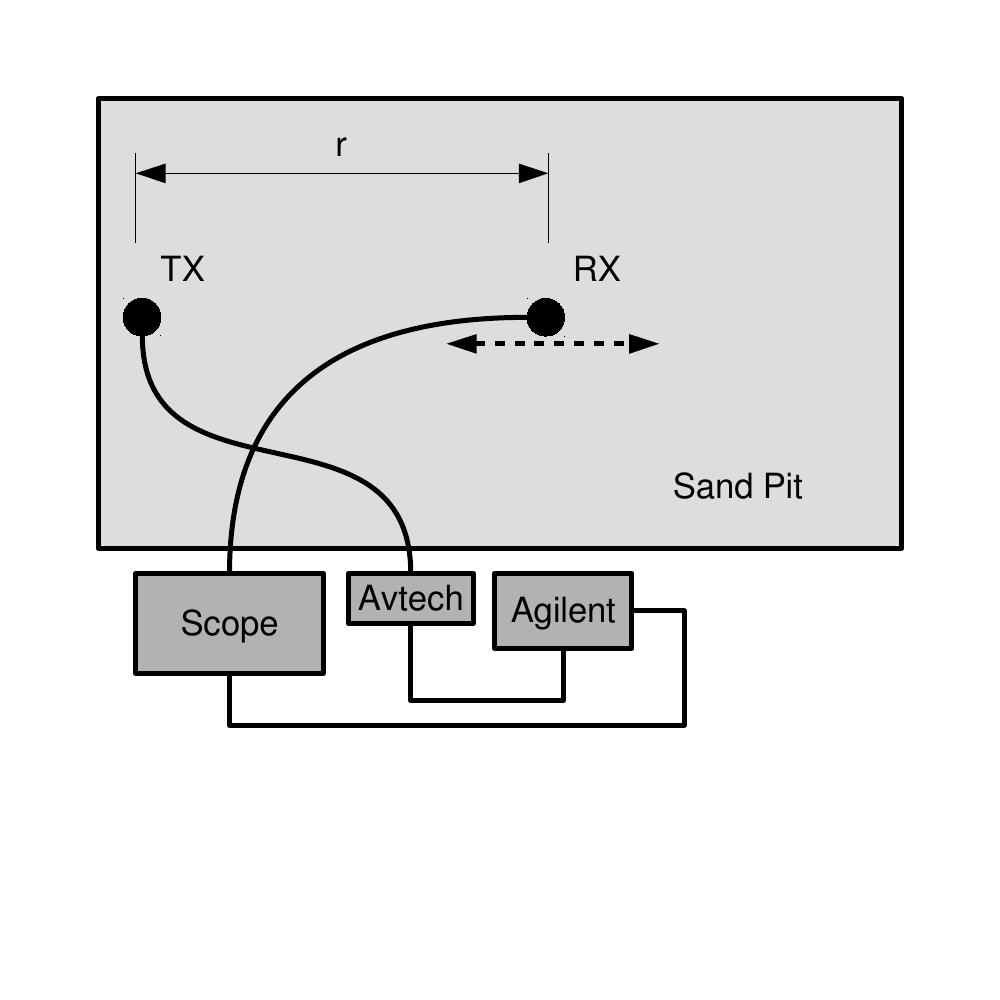}
	\caption{\it The outdoor sand pit experimental setup.  Signals from the receiver (Rx) were recorded on the 1 GHz oscilloscope at varying distances $r$ from the transmitter (Tx).  The Agilent pulse generator triggered the 400 ps-wide Avtech pulser, as well as the oscilloscope.\label{volleySet}}
	\end{center}
\end{figure}

Several field tests were performed to confirm the effects of the laboratory measurements.  Field data were recorded at an outdoor sand pit, where a RICE dipole receiver recorded waves produced by a RICE dipole transmitter, which radiated a 0.4 ns wide pulse from an Avtech AVL-200-C pulse generator.  The Avtech pulse generator, and the 1 GHz oscilloscope were triggered by an Agilent pulse generator.  The common triggering provided a stable time-base, meaning the relative timing of different receiver positions with respect to the transmitter could be compared.  Figure \ref{volleySet} shows a diagram of this setup.  The RICE dipoles respond well to broadband signals above 200 MHz, and the spectra extend to 500 MHz, the Nyquist limit of the oscilloscope in this case.  

\begin{figure}
\begin{minipage}{3.1in}
\begin{center}
	\includegraphics[width=8cm]{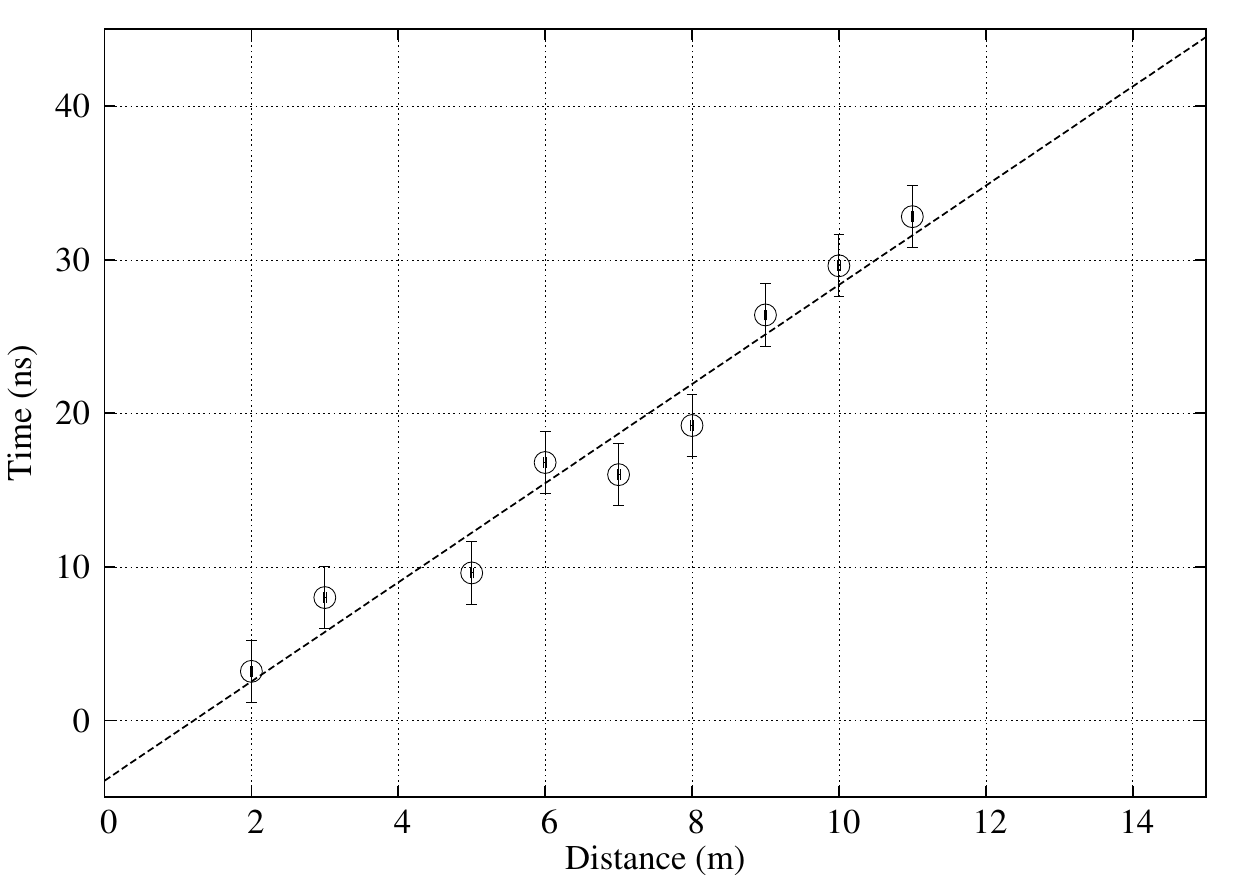}
	\caption{\it Cross-correlation times of the sand-pit surface setup (Fig. \ref{volleySet}), versus receiver position.  The $\chi^2/dof$ is 6.3/7.0, and the inverse of the fitted slope corresponds to the speed of the pulses in the system. The y-intercept is consistent with zero, within errors.\label{fit}}
	\end{center}
\end{minipage}
\hspace{0.3in}
\begin{minipage}{3.1in}
\begin{center}
	\includegraphics[width=8cm]{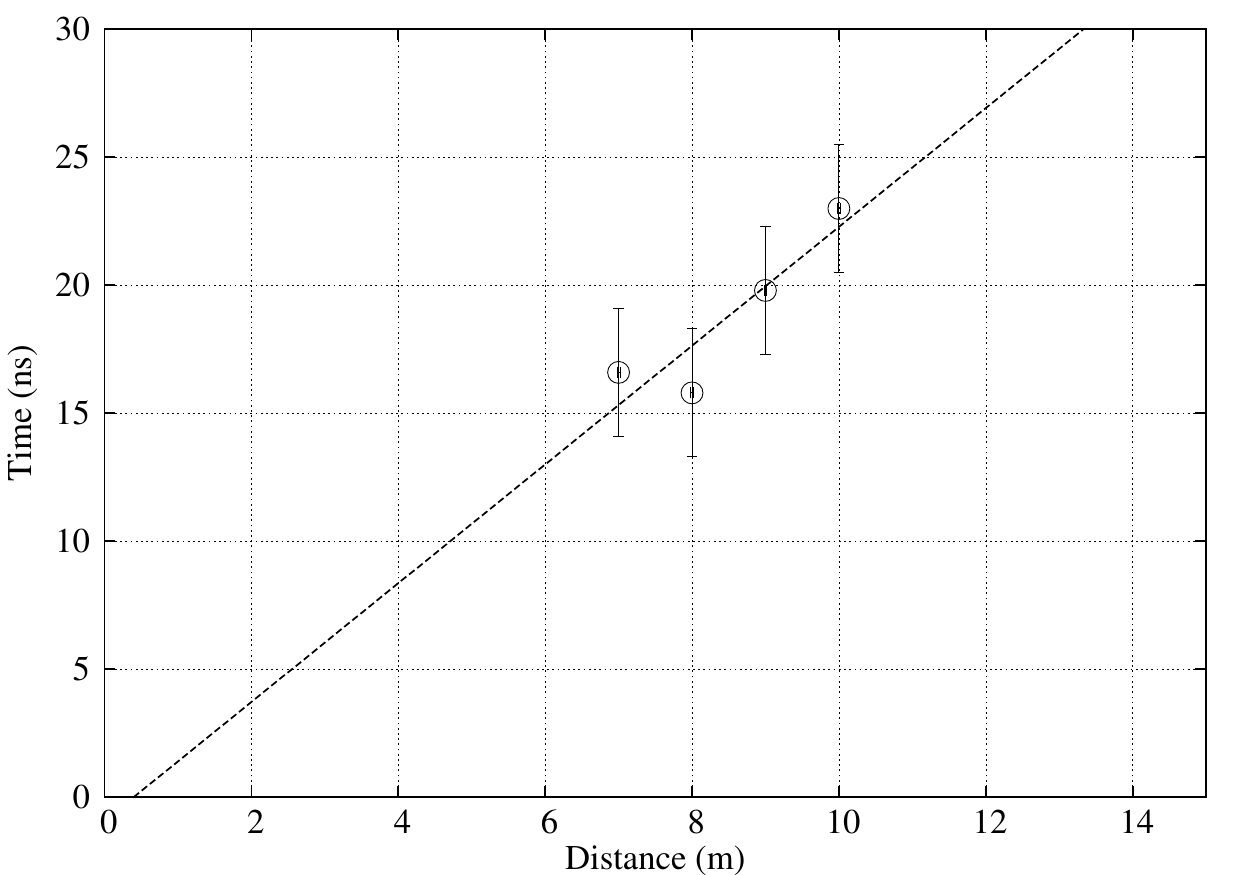}
	\caption{\it Cross-correlation times of the sand-pit surface setup (Fig. \ref{volleySet}), versus receiver position, with the transmitter and receiver half-buried.  $\chi^2/dof$=1.4/2.0; y-intercept is consistent with zero within errors.\label{fit2}}
	\end{center}
\end{minipage}
\end{figure}

Waveforms 
were recorded for separations between 2 and 12 m, for which the antennas were near the sand surface.  The time difference between the waveforms recorded at each distance relative to the first waveform (serving as a reference) is computed from the cross-correlation technique.  The cross-correlation of two strictly real signals is defined as in Eq. \ref{cross}.  The precision in the cross-correlation is 2.5 ns.  A linear fit was performed on the cross-correlation times versus the receiver distance.  In Fig. \ref{fit}, the x-axis gives the receiver position relative to the transmitter, and the y-axis contains the time lag that maximizes the cross-correlation between the receiver waveform and the reference waveform.

\begin{equation}
w_1 \star w_2 = \int_{-\infty}^{\infty} w_1(t) w_2(t+\tau) dt \label{cross}
\end{equation}

The fitted slope of the line yields a speed of $0.31\pm0.02$ m/ns, or $n=1.03\pm0.07$, consistent with through-air propagation.  The error on the distances in Fig. \ref{fit} is estimated at 2 cm.  The antennas were then buried half-way into the sand and the procedure was repeated.  
The corresponding speed (obtained from Fig. \ref{fit2}) is $0.4\pm0.1$ m/ns, or an implied index of $n = 1.3\pm0.3$.  The $\chi^2/dof$ for the half-buried case is $1.4/2.0$, corresponding to a confidence level of 70\%.  Although the index-of-refraction is less precise than the laboratory measurements, it is nevertheless consistent with those measurements.  The advantage of the outdoor sand-pit tests is that they probe lower frequencies, otherwise inaccessible in a smaller laboratory environment.  Table \ref{allSand} summarizes the average measured index-of-refraction values for surface setups at the various frequencies probed.

\begin{figure}
\begin{center}
	\includegraphics[width=10cm]{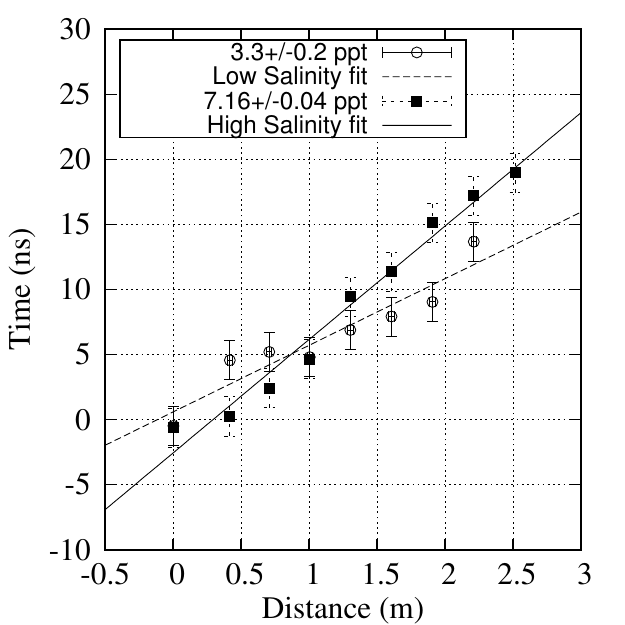}
	\caption{\it Propagation times of the salt-doped water surface setup (Fig. \ref{poolFig}), versus receiver separation.  The results are extracted from the slopes and summarized in Tab. \ref{waterTab}.  \label{waterSpeed}}
	\end{center}
\end{figure}

Direct speed measurements using the same technique were also performed for the salt-water pool setup.  Measurements were taken for two salinities: $3.3\pm0.2$ ppt and $7.16\pm0.04$ ppt.  The two salinity values were chosen to maximize both the salinity meter precision, and also to remain within the salinity meter measurement range.  
The distance between the two receivers was varied from 0.24-2.45 m (low salinity) vs. 0.24-2.75 m (high salinity).  
The signal arrival time is extracted from the time of the peak amplitude for the signals, and the speed is measured from the distance between the two receivers and the time difference of the peak amplitudes.  Despite the observed ringing and reflections, the uncertainty in the peak location was typically 1.5 ns.  Because the RICE dipoles were flush with the surface (well within 1$\lambda$), the difference between direct surface waves and waves reflecting from the surface at the specular point between the antennas is immeasurable.  However, waves traveling through air and reflecting from the surface would travel at $c_0$, rather than $c_0/n$.  Figure \ref{waterSpeed} demonstrates, however, that the speed of the waves in both the low and high salinity cases was less than $c_0$.  
The time between antenna channels is plotted versus channel separation, and the inverse of the slopes yields the speed.  The low salinity data corresponds to a wave speed $v = 0.20\pm0.03$ m/ns, and the high salinity data gives $v = 0.11\pm0.01$ m/ns, with the higher index-of-refraction corresponding to the higher salinity data.  The broadband measurements show some variation relative to the CW measurements at 2000 MHz, which indicated indices consistent with 1.0.  
Table \ref{waterTab} summarizes the fits, speed and index measurements.

We observe contrasting trends in the salinity and sand data.  The CW data originating from sand doped with metal shavings has significantly lower index than the other subsets of results.  The CW salt water and metal doped sand data show similar numbers for the index (near 1.0), indicating the geometry and conductivity of the setup plays a significant role.  However, the direct wave-speed measurements from broadband pulses yield $\emph{larger}$ numbers for the index.  This suggests there may be some frequency dependence in the index-of-refraction along the surface.  Both data and modeling above 1 GHz suggest that there is already dispersive behavior of the permittivity of water with dissolved ions\cite{Somaraju}.  The authors of that reference claim that the index is inversely proportional to salinity.  Our salinities are lower than those in Fig. 1 of\cite{Somaraju} by a factor of 3-10, but our index measurements are also lower.  We attribute this finding to the surface geometry, versus the bulk data used to create their Fig. 1.  However, since no significant super-luminal measurements are reported (other than the left-most point of Fig. \ref{hlamb}, which could be a chance coindidence) it is not likely that these are the classical surface waves sought for in our tests.  

\begin{table}
	\begin{center}
	\begin{tabular}{c c c c}
		\hline
		Salinity (ppt) & $\chi^2/dof$ & Speed (m/ns) & Index \\ \hline
		$3.3\pm0.2$ & 5.4/6 & $0.20\pm0.03$ & $1.5\pm0.2$ \\ \hline
		$7.16\pm0.04$ & 4.5/7 & $0.11\pm0.01$ & $2.7\pm0.2$ \\ \hline
	\end{tabular}
	\end{center}
	\caption{\it Summary of the pulse-timing results from the salt-doped water measurements.  The salinity was recorded in ppt by the salinity meter; the $\chi^2/dof$ of the linear fit of time vs. distance is given in the second column.  The speed and corresponding measured index-of-refraction are shown in the third and fourth columns, respectively. \label{waterTab}}
\end{table}

\subsection{Check of Inferred Index-of-Refraction from SWR Measurements}
\begin{figure}
\begin{minipage}{3.1in}
\begin{center}
	\includegraphics[width=8cm]{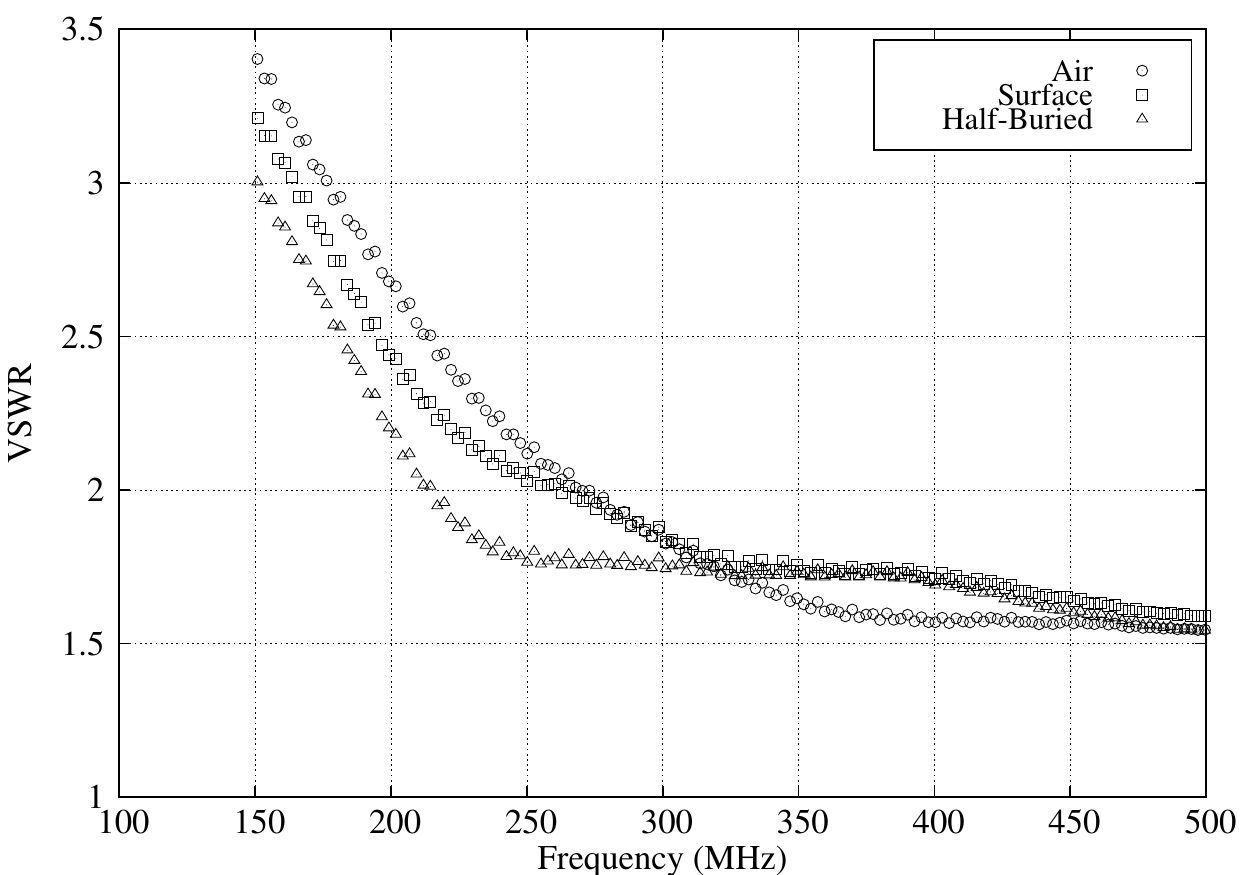}
	\caption{\it The VSWR parameter of the RICE dipole in air, near the sand surface, and half-buried in sand, taken at the outdoor sand pit.  The index-of-refraction for the near-surface and half-buried cases is derived from the low-frequency shift away from the air case (see text).  \label{vswr}}
	\end{center}
\end{minipage}
\hspace{0.3in}
\begin{minipage}{3.1in}
\begin{center}
	\includegraphics[width=8cm]{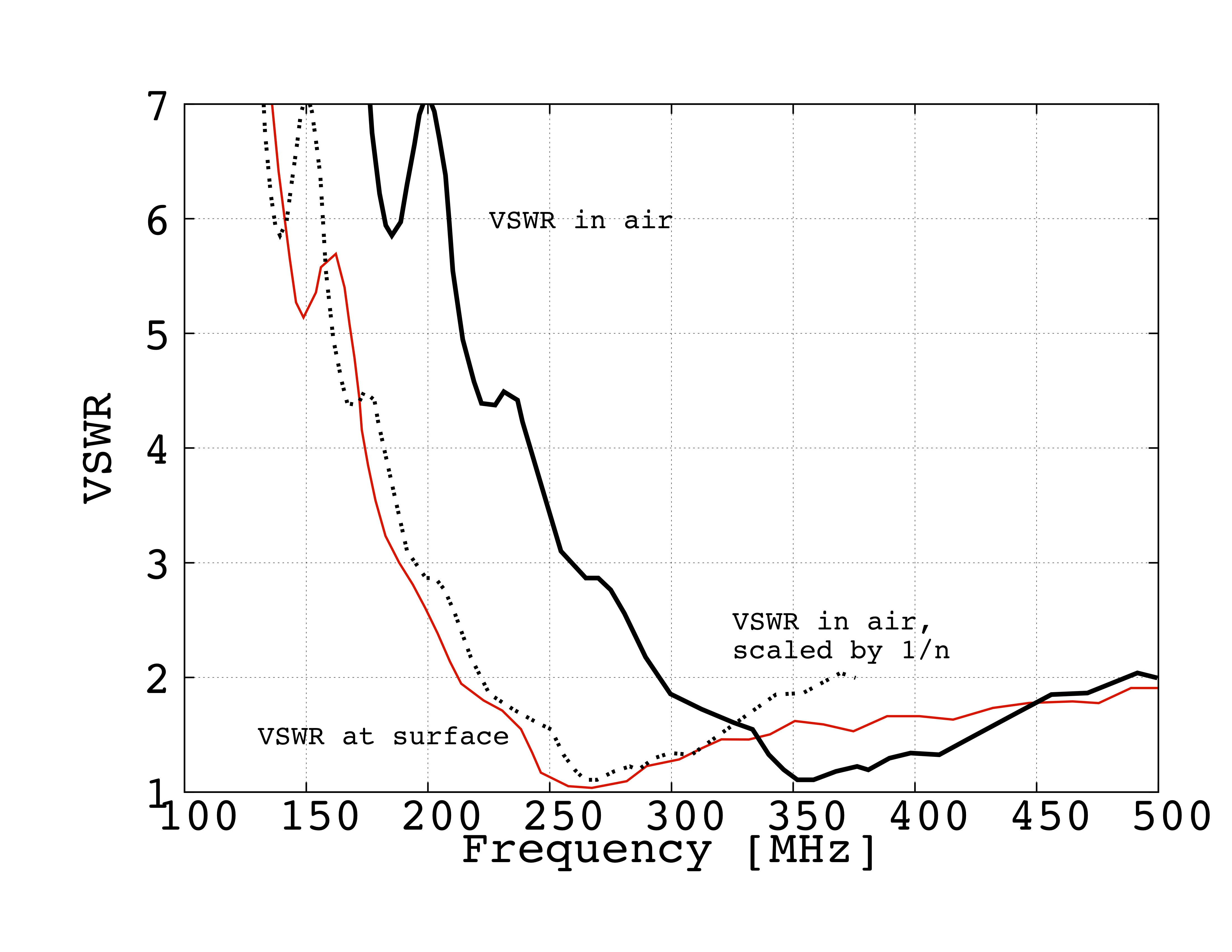}
	\caption{\it The VSWR parameter of the RICE dipole in air, half-way inside the salt surface in the Hutchinson salt mine pillar.  The index-of-refraction is derived in the same fashion as in Fig. \ref{vswr}.  \label{MarkJess}}
	\end{center}
\end{minipage}
\end{figure}

The VSWR was measured for the antenna at the transmitter location, for air, near surface, and half-buried positions.  In a dielectric with index $n$, the wavenumber scales like $k \rightarrow nk$.  Thus, the lowest frequency radiated efficiently by the antenna decreases by a factor of $n$. The VSWR parameter increases rapidly at low frequencies, which are inefficiently radiated by an antenna of limited size.  Due to the shift in the wavenumber in a dielectric, the location in frequency-space of this rapid increase in VSWR decreases by a factor of $n$.  We can therefore use the evolution of the VSWR with frequency as a measure of the index-of-refraction. The index was determined from the relative shifts in the VSWR spectra (shown in Fig. \ref{vswr}) for the cases of near-surface and half-buried antennas. The average measured refractive index for the near surface case is $n = 1.08\pm0.04$, and the half-buried measured refractive index is $n = 1.17\pm0.04$.  These results are in agreement with the results derived from the wave speed measurements, and are summarized in Tab. \ref{allSand} below.

Additional measurements were taken in a salt mine in Hutchinson, KS, where the smooth and extensive salt deposits provided a suitable dielectric surface.  The VSWR was measured in each case, confirming the expectation that near the surface, the low-frequency cutoff of antenna radiation decreased by a factor of $n$.  One such result is shown in Fig. \ref{MarkJess}.  The RICE dipoles were used as probes, where the relevant frequency is again $\sim$150 MHz (the cut-in frequency of the dipoles near the surface).  The antennas were placed perpendicular to the vertical salt wall; i.e., horizontal with respect to the ground.  The height of the antennas above the ground was approximately 2 m.  The shift required to overlay the air and in-salt spectra corresponds to $n=1.35\pm0.1$, which is in agreement with results obtained using the phased technique in the laboratory salt setup.

\subsection{Summary of Index-of-Refraction Results}

Tables \ref{allSand} and \ref{allSalt} summarize the techniques and materials used in the surface index-of-refraction measurements involving pure sand and salt.  Table \ref{lastTab} sumarizes the results of the salt water surface measurements. In general, our data are consistent with the expectation that, within approximately one wavelength of a dielectric with $n_{dielectric}>$1, the measured index-of-refraction is intermediate between that of air and that of the pure dielectric. The refractive index of bulk sand is approximately 1.6\cite{dry}, however our measurements of the surface index of sand indicate lower values.  Similarly, NaCl has an index of 2.4-2.8 in the bulk, however, our near-surface measurements cluster around $n\approx 1.5$.  For the `true' sand index value, data are sparse at radio wavelengths, however it can be extrapolated from the optical regime, as it is essentially flat for $\lambda>$0.1 cm. 
The `true' value for bulk salt derives from two sources\cite{salt1,salt2}, including studies investigating the potential for large rock-salt formations as neutrino detectors. 

\begin{table}
	\begin{center}
	\begin{tabular}{c c c}
		\hline
		Frequency (MHz) & Setup & index near surface \\ \hline
		1000 & Lab, small dipoles & $1.22\pm0.03$ \\ \hline
		1500 & Lab, small dipoles & $1.20\pm0.03$ \\ \hline
		150-500 & RICE dipoles near surface (wave speed) & $1.03\pm0.07$ \\ \hline
		150-500 & RICE dipoles half-buried (wave speed) & $1.3\pm0.3$ \\ \hline
		150-500 & RICE dipoles near surface (VSWR) & $1.08\pm0.02$ \\ \hline
		150-500 & RICE dipoles half-buried (VSWR) & $1.17\pm0.02$ \\ \hline
	\end{tabular}
	\end{center}
	\caption{\it A summary of all sand surface index-of-refraction measurements. The first two rows are obtained from the phased technique in the laboratory at two different frequencies.  The second two rows are obtained from outdoor measurements via cross-correlation techniques.  The third pair of rows are obtained from outdoor VSWR measurements. In this final pair, the frequency of the VSWR minimum is chosen as the representative frequency. The wave speed measurements are broadband; the phased technique is performed at constant frequency.\label{allSand}}
\end{table}

\begin{table}
	\begin{center}
	\begin{tabular}{c c c}
		\hline
		Frequency (MHz) & Setup & index near surface \\ \hline
		1000 & Lab, small dipoles & $1.50\pm0.01$ \\ \hline
		1500 & Lab, small dipoles & $1.45\pm0.01$ \\ \hline
		150-500 & RICE dipoles near salt-pillar (VSWR) & $1.35\pm0.1$ \\ \hline
	\end{tabular}
	\end{center}
	\caption{\it A summary of all salt surface index-of-refraction measurements. The first two rows are obtained from the phased technique in the laboratory at two different frequencies, and the third row is obtained from a salt pillar in a salt mine. \label{allSalt}}
\end{table}

\begin{table}
	\begin{center}
	\begin{tabular}{c c c}
		\hline
		Frequency (MHz) & Setup & index near surface \\ \hline
		150-500 & RICE dipoles near surface (wave speed) & $1.5\pm0.2$ ($3.3\pm0.2$ ppt) \\ \hline
		150-500 & RICE dipoles near surface (wave speed) & $2.7\pm0.2$ ($7.16\pm0.04$ ppt) \\ \hline
		2000 & Small dipoles w/ Quad horn attachment & $0.95\pm0.1$ ($h/\lambda=2.94$) \\ \hline
		2000 & Small dipoles w/ Quad horn attachment & $0.95\pm0.06$ ($h/\lambda=2.00$) \\ \hline
		2000 & Small dipoles w/ Quad horn attachment & $0.97\pm0.08$ ($h/\lambda=1.33$) \\ \hline
		2000 & Small dipoles w/ Quad horn attachment & $0.85\pm0.10$ ($h/\lambda=1.00$) \\ \hline
	\end{tabular}
	\end{center}
	\caption{\it A summary of the salt water measurements.  The first two rows correspond to wave-speed measurements made with RICE dipole antennas, and pulse-timing, at two different salinities.  The final four rows show the results of the phased technique, where the ratio $h/\lambda$ was varied from 2.94 to 1.00 ($h$ is the antenna height above the surface, and $\lambda$ is the free space wavelength).  The salinity for the final four rows was $7.16\pm0.04$ ppt in added salt. \label{lastTab}}
\end{table}

The measurements are self-consistent, with the possible exception of the phased technique near the water surface.  The sand measurements show that much larger RICE dipoles produce similar results as simple, lab-built 8 cm dipoles attached to coaxial cables, for three different techniques.  The salt water experiments produce a similar effect when the speed is recorded from pulse timing, in that the index results are greater than 1.0 but lower than that of salt water in bulk. 
Although the statistics are too limited to draw definitive conclusions, we do observe an interesting trend in the salt-water measurements, which, for f=2000 MHz, consistently yields $n<$1, with the deviation from 1 largest for the case where the antennas are closest to the salt-water surface. The overall combined significance of the deviation from 1 is equal to 1.94$\sigma$. Of course, phase velocities exceeding $c_0$ in dispersive media are well-known; given somewhat larger scale laboratory test dielectrics, a successor experiment might probe the frequency dependence over the full range from 1--10000 MHz and map out dispersive effects somewhat more comprehensively.


\section{Summary and Conclusions}
Using a wide range of dielectrics, including pure sand, a sand+metal mixture, ice, and water with varying amounts of saline impurities, and over a range of frequencies spanning 100 MHz--4000 MHz, we have searched for unambiguous evidence for surface wave coupling. Each individual experiment includes its own particular systematic errors, which we have not fully explored, but rather implicitly ``average'' over the variety of systematics inherent in our measurements. We find no such compelling evidence, and conclude that surface radio-frequency wave detection of neutrinos is unlikely to be a promising experimental avenue.

\section{Acknowledgements}
Data collection for this experiment was primarily performed by Kansas high school students during the summers of 2012, 2013 and 2014; correspondingly, 
we would like to thank the Fermi National Laboratory and the Department of Energy for sponsoring the QuarkNet program, through which American high-school physics students are connected with University researchers.  We also would like to acknowledge Marie Piasecki (NASA) and John Ralston (KU) for their helpful discussions early in this project, and Jeffrey Engelstad for his assistance with the Great Sand Dunes National Park measurements.  Finally, we thank the staff at KU Facilities Management and the Hutchinson Salt Mine for their support and cooperation.

\bibliography{SW15.bib}
\bibliographystyle{unsrt}

\end{document}